\documentclass[aps,twocolumn,superscriptaddress,amssymb,notitlepage, nofootinbib]{revtex4-1}

\usepackage[utf8]{inputenc}
\usepackage[english]{babel}
\usepackage[T1]{fontenc}
\usepackage{hyperref}

\usepackage{amsfonts,amssymb,amsmath} 
\usepackage{graphicx}
\usepackage{soul}

\graphicspath{{./figs/}}

\begin{abstract}
The evaluation of the Jones polynomial at roots of unity is a paradigmatic problem for quantum computers.
In this work we present experimental results obtained from existing noisy quantum computers for special cases of this problem, where it is classically tractable.
Our approach relies on the reduction of the problem of evaluating the Jones polynomial of a knot at lattice roots of unity to the problem of computing quantum amplitudes of qudit stabiliser circuits,
which are classically efficiently simulatable.
More specifically, we focus on evaluation at the fourth root of unity, which is a lattice root of unity, where
the problem reduces to evaluating amplitudes of qubit stabiliser circuits.
To estimate the real and imaginary parts of the amplitudes up to additive error we use the Hadamard test, yielding non-Clifford circuits that nevertheless we can always efficiently compute the correct output of.
Hence, we further argue that this setup defines a standard benchmark for near-term noisy quantum processors.
Additionally, we study the benefit of performing quantum error mitigation with the method of zero noise extrapolation. 
\end{abstract}

\begin{document}

\title{Estimating the Jones polynomial for Ising anyons on noisy quantum computers}

\author{Chris N. Self}
\email{christopher.self@quantinuum.com}
\affiliation{QOLS, Blackett Laboratory, Imperial College London SW7 2AZ, United Kingdom}
\affiliation{Quantinuum, Partnership House, Carlisle Place, London SW1P 1BX, United Kingdom}

\author{Sofyan Iblisdir}
\affiliation{Dpto. An\'alisis Matem\'atico y Matem\'atica Aplicada, Facultad de Matem\'aticas, Universidad Complutense, 28040 Madrid, Spain}
\affiliation{Dpt. Astrof\'isica i F\'isica Qu\`antica \& Institut de Ci\`encies del Cosmos, Facultat de F\'isica, Universitat de Barcelona, 08028 Barcelona, Spain}

\author{Gavin K. Brennen}
\affiliation{Center for Engineered Quantum Systems, Department of Physics \& Astronomy, Macquarie University, 2109 NSW, Australia}

\author{Konstantinos Meichanetzidis}
\email{k.mei@quantinuum.com}
\affiliation{Department of Computer Science, University of Oxford}
\affiliation{Quantinuum, 17 Beaumont St., OX1 2NA, Oxford, United Kingdom}

\date{June 16, 2025}

\maketitle

\section{Introduction}

Knot theory is of both theoretical and practical interest to a wide range of areas of research \cite{Adams2004,Kauffman2001}.
A fundamental question of knot theory is distinguishing when two knot representations, or more generally links which are multicomponent knots, are topologically equivalent.
This is addressed by the notion of the link invariant,
a mathematical quantity extractable from a link
which is independent of the representation of the link.
A famous link invariant is the Jones polynomial,
a polynomial in one complex variable.

The evaluation of the Jones polynomial of a link at roots of unity
is important for the field of quantum computation.
Any other problem efficiently computable on a quantum computer
can be reduced to it.
Specifically, the problem reduces to estimating a quantum amplitude, involving a quantum circuit dictated by the link, up to additive error \cite{Aharonov2009}.
The quantum protocol used for this is the Hadamard test (H-test).

Currently available quantum devices fall under the noisy intermediate-scale quantum (NISQ) paradigm;
they are not error-corrected and so the size of circuits one can run is limited by the amount of decoherence.
To avoid the detrimental effects of decoherence, it is crucial to adapt abstract circuits to the specific quantum processor and reduce the number of operations required, one aims to optimise compilation of abstract circuits to circuits composed of the native gateset of the specific quantum processor, and also respect the processors qubit-qubit connectivity.
Furthermore, there is a plethora of error mitigation techniques being developed~\cite{endo2021hybrid}, whose aim is to amplify and correct the biases of the signal that one reads-off a quantum device.

In this work, we are concerned with two theoretical reductions,
and we focus on knots, however our pipeline is readily applicable to links, as well.
The first reduction maps the evaluation of the Jones polynomial of a knot, to
the computation of the partition function of an associated Potts model.
The second reduction maps the latter to the estimation of a quantum amplitude involving a stabiliser, or Clifford, unitary. 
Clifford unitaries are classically efficiently strongly simulatable, where strong simulation means that any complex amplitude can be obtained exactly, so there is no quantum advantage for this task. 
However, the quantum circuits for estimating the amplitudes through the H-test are generally not Clifford circuits since they involve the controlled unitary. 
This is a useful situation for application benchmarking~\cite{lubinski2023application, QCAppBenchmarks}, since the circuits to be executed are made from universal gates but the expected answer can be efficiently classically obtained even for large numbers of qubits.
Hence we propose that sampling knots and estimating the Jones polynomial on a quantum device while verifying the answer efficiently by classical simulation can be automated, and defines a standardised application benchmark for quantum computers.
We focus on a small illustrative example of the procedure, and discuss the quantum engineering aspect of compiling the circuit for the H-test on IBM Quantum backends and performing error mitigation with comparison of different methods.
Furthermore, we track the performance over days and check consistency of the devices' operation.

\section{Knot diagrams and the Jones polynomial}
\label{sec:knot-diagrams}

A \emph{knot} is a circle embedded in $\mathbb{R}^3$.
Intuitively, it is a strand that can be tangled with itself and is closed, i.e. has no open, or loose, ends.
Importantly, strands are not allowed to intersect nor go through themselves or other strands.
A \emph{knot diagram} is a projection of a knot in $\mathbb{R}^2$ which keeps information about over- and under-crossings.
The following are examples of knot diagrams \cite{knotatlas}:
\begin{center}
\includegraphics[width=0.7\columnwidth]{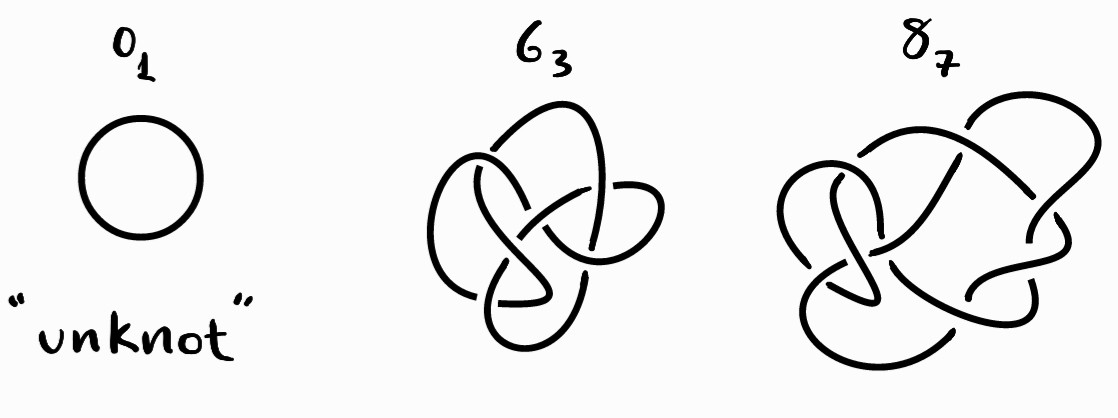}
\end{center}

Different projections of the \emph{same} knot,
result in diagrams that can be \emph{smoothly deformed to each other}.
In terms of diagrams, these {smooth} deformations are generated by the three Reidemeister moves:
\begin{center}
\includegraphics[width=0.4\columnwidth]{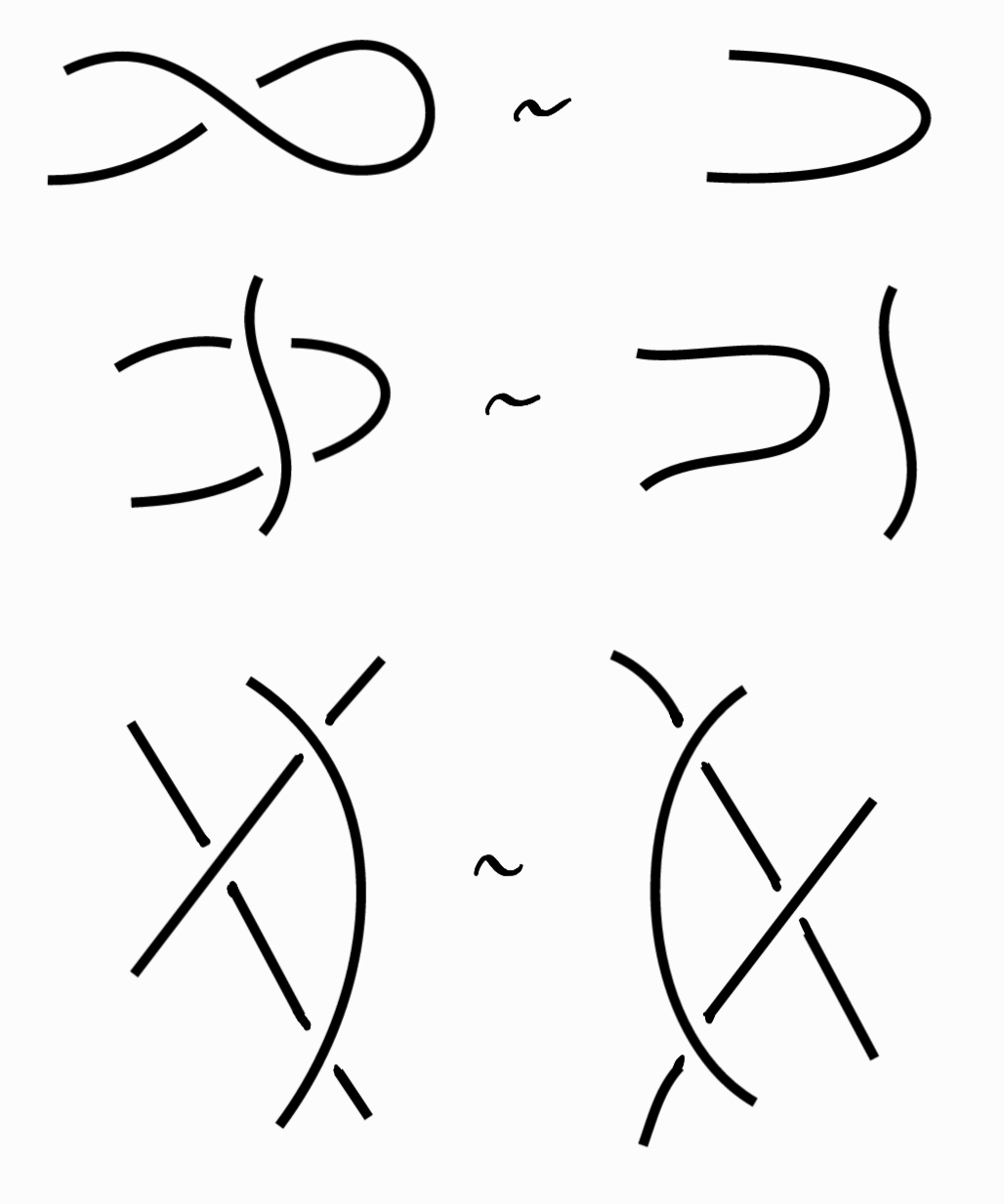}
\end{center}

The Jones polynomial $V_K(t)$ of a knot $K$ is a Laurent polynomial in a complex variable $t\in\mathbb{C}$ and it can be obtained by a diagram of $K$.
It serves as a knot \emph{invariant}
in the sense that $K \sim K' \Rightarrow V_K = V_{K'}$.
By $K \sim K'$ we denote topological equivalence
in the sense that the diagram of $K$ can be rewritten to the diagram of $K'$ by a \emph{sequence of Reidemeister moves}.
In other words, two diagrams $K$, $K'$ are topologically equivalent if $K$ can be smoothly deformed to $K'$ by moving, bending and wiggling the strands, but without cutting or gluing them.
However, note that $V_K$ is \emph{not} a complete invariant,
as in general $V_K = V_{K'} \nRightarrow K \sim K'$.
The Jones polynomial is normalised such that $V_\text{unknot}=1$.
Interestingly, it is conjectured that the unknot is the only knot for which this is true.

An elegant way of computing the Jones polynomial from a knot diagram is via the Kauffman bracket \cite{Kauffman2001},
which involves breaking every crossing into two types of avoided crossings, incurring {a number of operations \textit{exponential} in the number of crossings.} 
In particular, the Jones polynomial is an invariant of \emph{oriented} knots, as it is sensitive to the handedness or the \emph{writhe}, $w$.
On an oriented diagram, a {+1 value} is assigned to a right-handed crossing, a {-1 value} to a left-handed one, and their sum is $w$.
For example:
\begin{center}
\includegraphics[width=0.55\columnwidth]{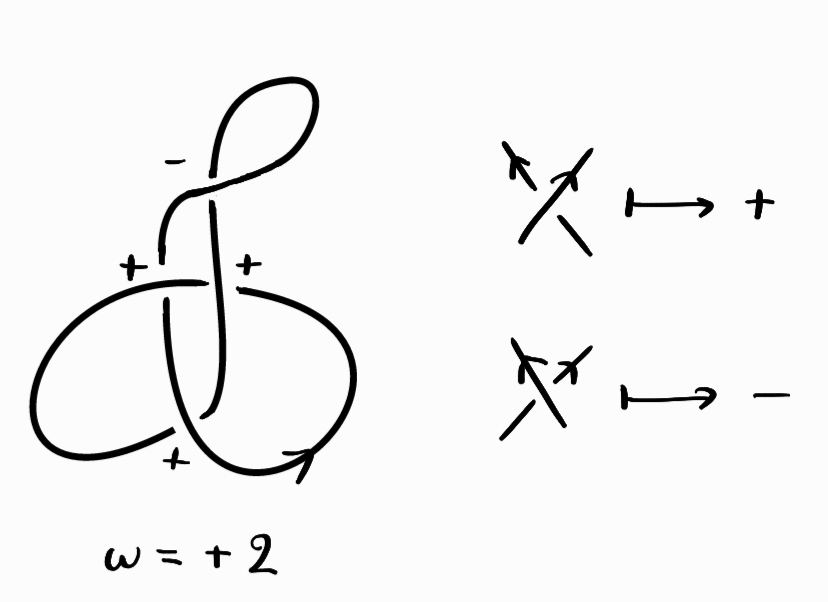}
\end{center}
The writhe can be computed efficiently,
and multiplying the Kauffman bracket with a prefactor depending on $w$
results in the Jones polynomial.
In fact, the maximum and minimum degrees of the polynomial can be inferred from the knot diagram efficiently and are upper bounded by a linear function in the number of crossings.
The exponential cost needed to compute the polynomial is reflected in that {its} coefficients can get exponentially large.

\section{ Jones polynomial evaluation via Potts partition function}

It is in general \#P-hard to \emph{evaluate} the polynomial at a particular value $t$ on the complex plane. For some values of $t$, the
problem can be reduced to the computation of the partition functions of a particular Potts model as follows \cite{Wu}.

From the diagram of a knot $K$ we extract the signed graph $G_K$, called the \emph{Tait graph}, by bicolouring the diagram into black and white regions, with the convention that the background is white and colours on either side of each strand are different. This "checkerboard" bicolouring is always possible as the crossings are all four-valent. We assign vertices to the black regions and edges that connect the black regions through crossings. The edges are assigned signs, called the Tait signs, according to a rule that depends on the crossing being over or under and also the colours around the crossing. The sum of the Tait signs is the Tait number $\tau$, and it can be obtained efficiently. An example of this procedure is illustrated as follows:
\begin{center}
\includegraphics[width=0.8\columnwidth]{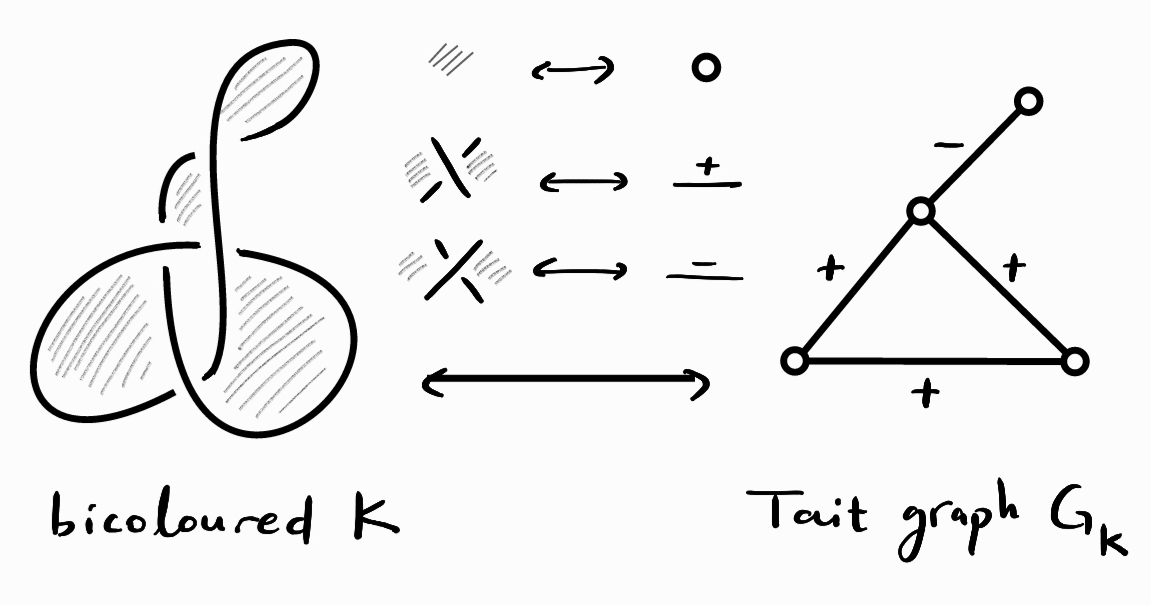}
\end{center}
Note, the crossing signs, which sum to $w$, and the Tait signs, which sum to $\tau$, should not be confused, as they are different in general.

To arrive at a Potts model,
we place $q$-state classical spins $\sigma_i$ on the vertices of the obtained graph
and have them interact \emph{along the edges}.
Spins interact only when they are in the same state and the coupling $J_{ij}\in\{J_+, J_-\}$ depends on the Tait-sign of the corresponding edge $(i,j)$.
The couplings are related to the Jones variable $t$ by
$e^{J_\pm}=-t^{\mp 1}$,
which in turn relates to the spin dimension as 
$q=t+t^{-1}+2$.
Solving for $t$ we obtain
\begin{equation}\label{eq:eval-point}
t(q)=\frac{1}{2}(q+\sqrt{q}\sqrt{q-4} -2),
\end{equation}
which tells us at which point $t\in\mathbb{C}$ we evaluate the Jones polynomial depending on our choice of the dimension $q\in\mathbb{N}$ of the spins.
The spin-spin interactions are designed such that
the partition function of this Potts model,
\begin{equation}\label{eq:part-func}
\mathcal{Z}_{K}(q) = \sum_{\{\sigma\}} \prod_{(i,j)} e^{\delta_{\sigma_i \sigma_j} J_{ij} } \in \mathbb{C} ,
\end{equation}
is a knot invariant under the Reidemeister moves, which now correspond to graph operations:
\begin{center}
\includegraphics[width=0.75\columnwidth]{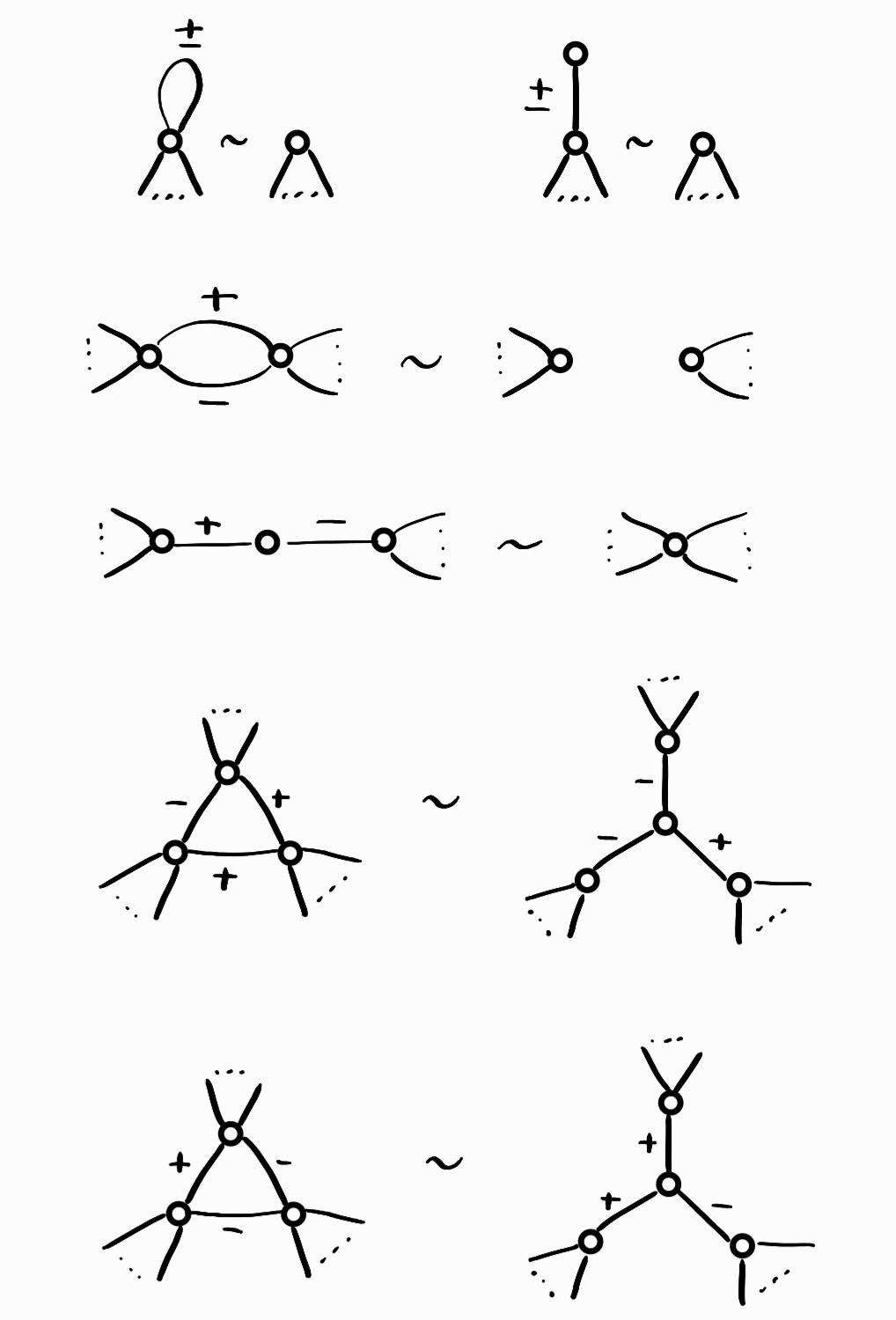}
\end{center}

Then, by this construction, the
Jones polynomial is related to this Potts model's partition function as
$$V_K(t(q)) = \mathcal{A}(t(q),\tau,w,n) \mathcal{Z}_{K}(q).$$
The proportionality factor
$$\mathcal{A} = (-t^{\frac{1}{2}}-t^{-\frac{1}{2}})^{-(n+1)} (-t^{{3}/{4}})^w t^{{\tau}/{4}}$$
depends on the evaluation point $t(q)$, the number of spins $n$, the writhe of the knot $w$, and the Tait number $\tau$ of the Tait graph, so it can be computed efficiently.
The computationally expensive quantity {here} is the partition function {$\mathcal{Z}_{K}(q)$ }, the exact computation of which, is a \#P-hard problem in general.

\section{Potts partition function as a tensor network}

For $q\in\mathbb{N}$,
the spins can occupy the states $\sigma_i\in\{0,1,\dots,q-1\}$.
In this case, the scalar quantity $\mathcal{Z}_K(q)$ of Eq.~\ref{eq:part-func} can be represented as a tensor network $T_K$ with $G_K$ the underlying graph and
$q$ the bond-dimension \cite{Meichanetzidis2019}.
Every vertex is assigned a `spider' tensor and every signed edge is assigned a sign-dependent matrix.
For example:
\begin{center}
\includegraphics[width=0.7\columnwidth]{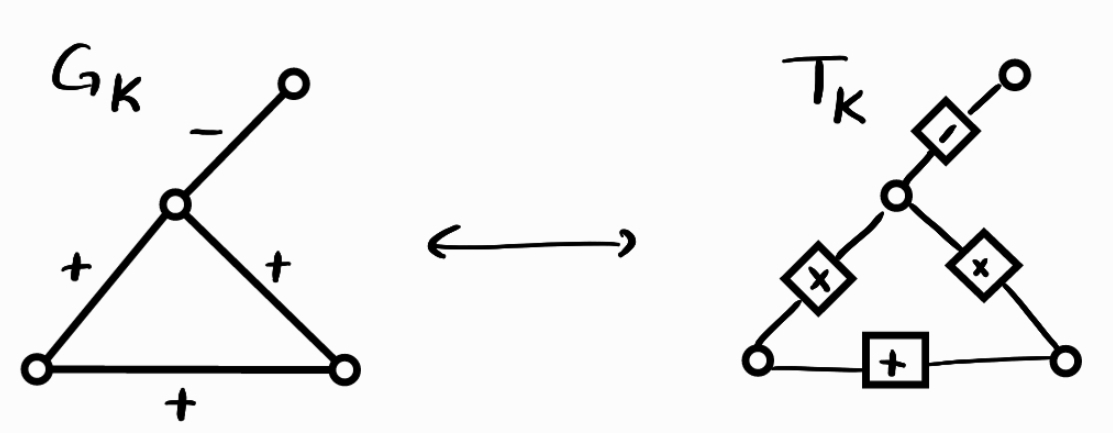}
\end{center}
where the 'spider' tensors, or 'copy' tensors, and the $\pm$-matrices are defined as follows:
\begin{center}
\includegraphics[width=0.85\columnwidth]{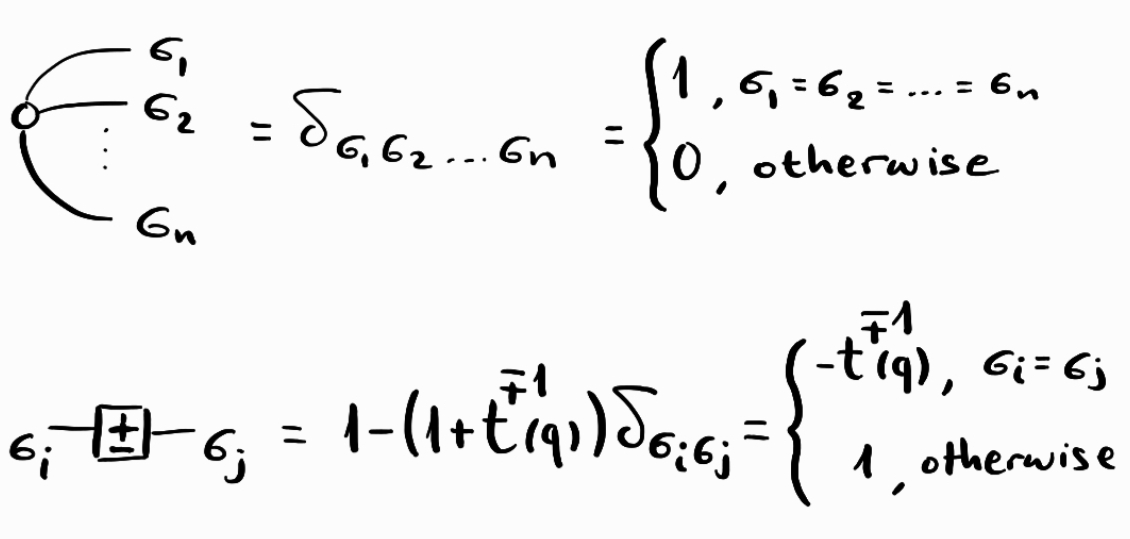}
\end{center}

The $\pm$-matrices encode the contributions to the partition function due to the spin-spin interactions.
The spiders represent hyperindices, which when summed over realise the sum over all possible spin configurations.
This sum is equivalent to the full contraction of $T_K$,
i.e. performing all sums over common indices represented by the edges in the tensor network.
Full tensor contraction of $T_K$ then returns $\mathcal{Z}_K(q)$.
Note that full contraction of generic tensor networks is a \#P-hard problem \cite{Damm2002}.
Interestingly,
tensor contraction algorithms based on graph-partitioning
exhibit subexponential space and time complexity, $O(q^{\sqrt{n}})$, where $n$ is the number of tensors in the network, when the underlying graph is planar \cite{Gray2021}. This is indeed the case for tensor networks obtained from Tait graphs.
In Ref. \cite{Meichanetzidis2019}, such methods were employed to exactly compute $\mathcal{Z}_{K}(q)$ for any $q\in\mathbb{N}$
where we refer the reader for more detail.

However, it is known that
at the `lattice roots of unity',
$\theta_\Lambda = \{e^{i\theta} |~ \theta=\pm \pi , \pm \frac{\pi}{2}, \pm \frac{\pi}{3}, \pm \frac{2\pi}{3} \}$,
the problem is in P \cite{jaegervertiganwelsh1990}.
For the specific values of integer spin dimension we have that
$q\in\{2,3,4\} \Leftrightarrow t(q) \in \theta_\Lambda$,
which means that at these points the computation of $\mathcal{Z}_K(q)$
is in P.
Using the ZX-calculus \cite{coecke_kissinger_2017,jvdwZX},
Ref. \cite{Teague} presents efficient simplification stategies for these specific tensor networks, using rewrite rules of the qubit and qutrit ZX-calculus, providing an alternative proof of the tractability of evaluating the Jones polynomial at lattice roots of unity.
We would like to stress here that the use of formal graphical languages help bridge existing algorithms for the Jones polynomial to currently available quantum technology, as we will see in the next section.
We refer the reader to the work cited above for more detailed expositions.

\section{From tensor network to quantum circuit}
\label{sec:tensor-nets-to-qcircs}

For $q\in\{2,3,4\}$, the tensor network $T_K$ provides the blueprint for a quantum computation in the form of an $n$-qudit quantum circuit $C_K$ with
specific input states and postselections,
representing the quantum amplitude $\mathcal{Z}_{K}(q) = {}^{\otimes n}\langle+| C_{K}(q) |+\rangle^{\otimes n} \in\mathbb{C}$ \cite{Iblisdir2014},
where $|+\rangle = \sum_{i=0}^{q-1}|i\rangle$ is the unnormalised plus state.
To make this apparent, we use the `fusion' rule obeyed by the spiders \cite{jvdwZX}, according to which any two spiders connected by at least one `wire', i.e. a common index, they can be fused to a single spider which inherits the open wires of the two fused spiders:
\begin{center}
\includegraphics[width=0.4\columnwidth]{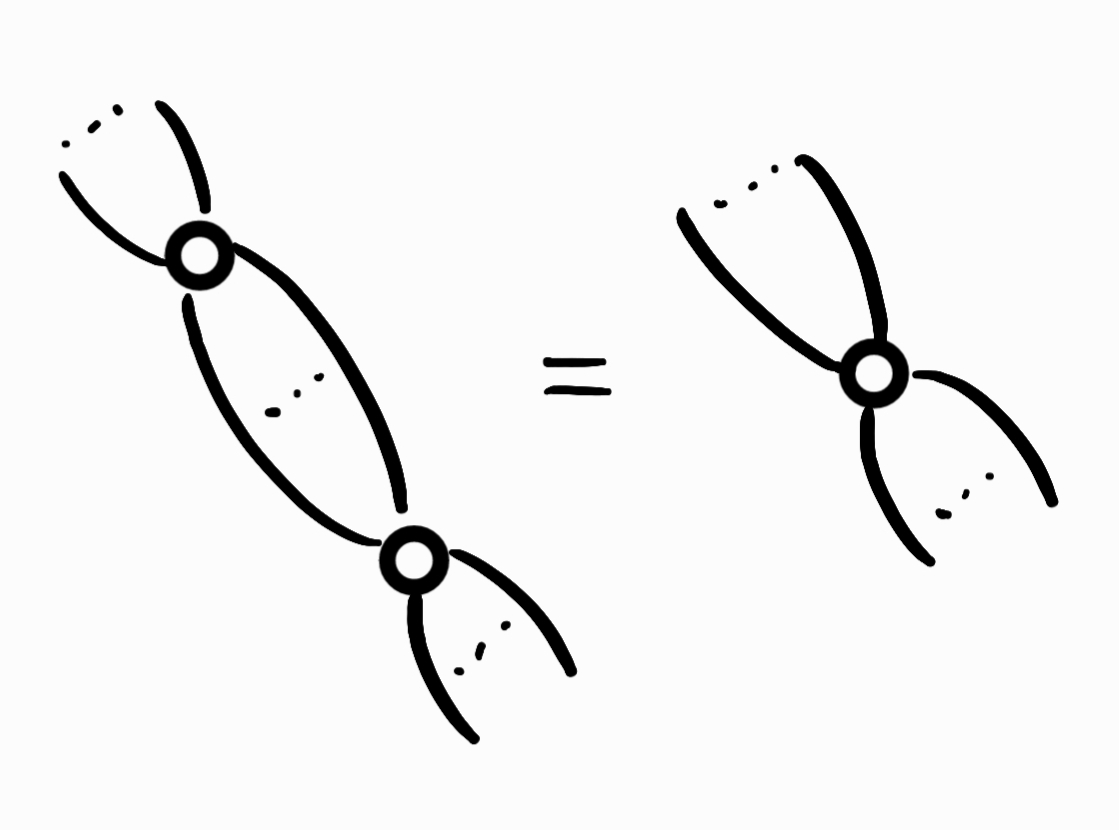}
\end{center}
The fusion rule immediately follows from the fact that the spider
acts as a copy operation on basis states of the vector spaces defined on its `wires', and can be verified by performing the explicit tensor contraction along the common indices and the properties of the Kronecker delta in terms of which the spider is defined.
The fusion rule then allows us to `pull out' the input and output plus-states and interpret the tensor network as a quantum circuit. For example:
\begin{center}
\includegraphics[width=0.8\columnwidth]{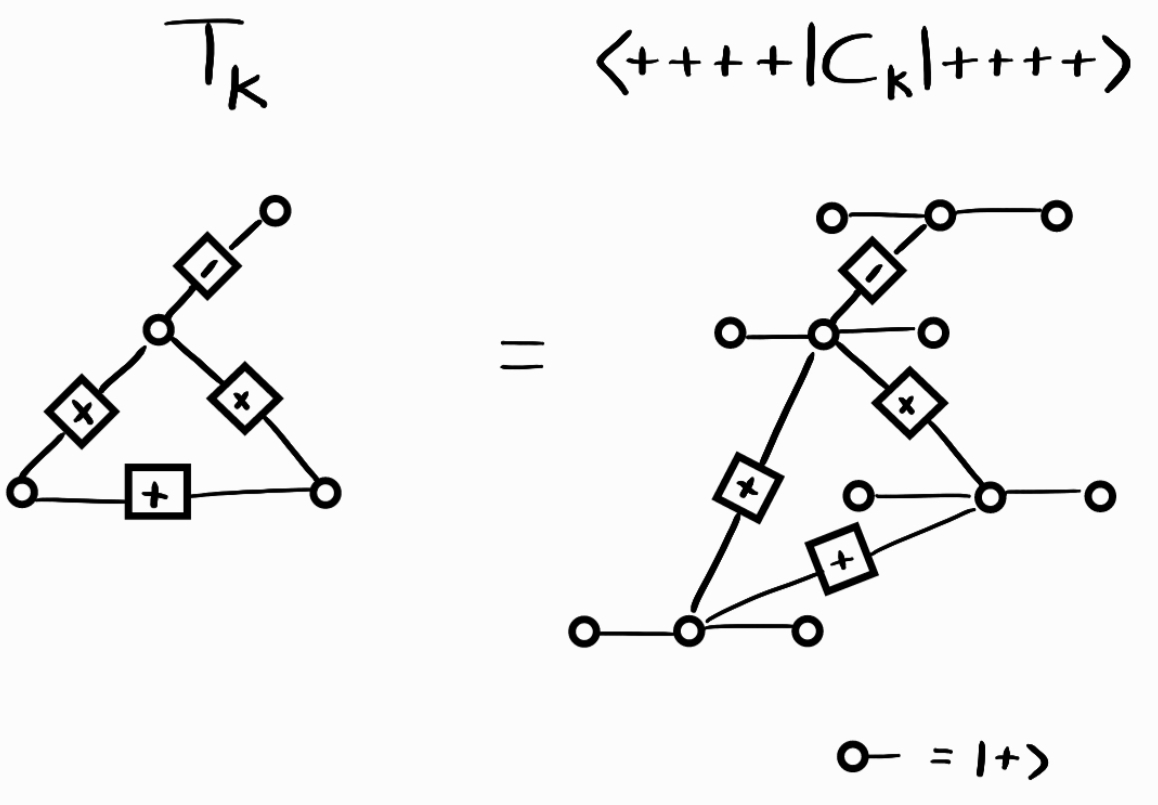}
\end{center}
Reading-off $C_K$ from $T_K$ is straightforward.
Every edge that goes through a $\pm$-matrix can be interpreted as a
unitary gate $\mathcal{K}^\pm$ acting on two $q$-level qudits:
\begin{center}
\includegraphics[width=0.8\columnwidth]{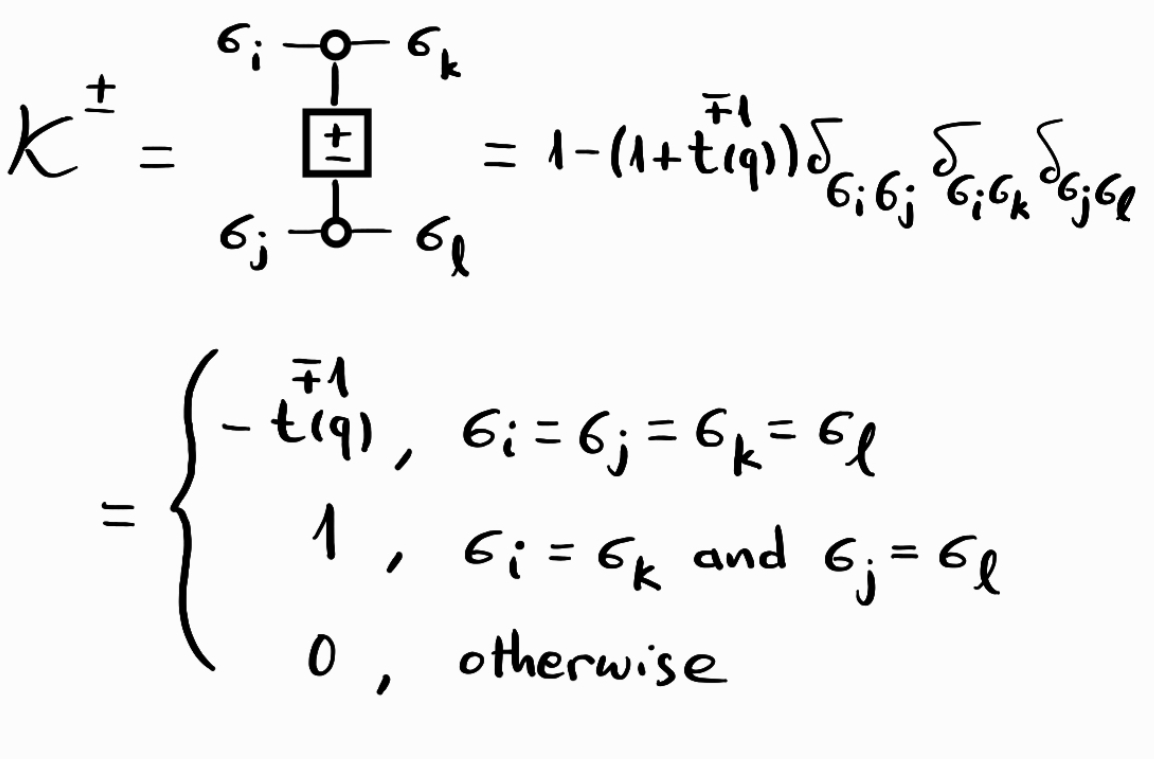}
\end{center}
These gates are diagonal and so they commute,
which is also seen by the fusion rule which allows us to draw them all in one layer.
Thus $T_K$ defines a special case of an {instantaneous quantum polynomial} (IQP) circuit, simply described by the edge list of $G_{K}$;
the edges can be read-off in any order and the edge's Tait sign indicates whether to perform a $\mathcal{K}^+$ or a $\mathcal{K}^-$ gate.

Evaluating $V_K(t)$ at generic roots of unity $t=e^{i\theta}$
is a paradigmatic BQP-complete problem.
However, here we are working in a setting where these amplitudes correspond to the evaluation of the Jones polynomial at lattice roots of unity.
The tractability of the problem for these cases manifests in $C_{K}$
being a \emph{Clifford} circuit,
defined on qubits for $q\in\{2,4\}$ and qutrits for $q=3$.
Such circuits can be simulated \emph{classically efficiently}, as stated by the Gottesman-Knill theorem \cite{gottesman}.
The results of Refs. \cite{Duncan2020,Teague} can also be understood as a graphical version of this result given in terms of diagrammatic rewrite strategies.

Note that the $\mathcal{K}^\pm$-gates are not unitary for integers $q\geq 5$, since then $t(q)$ is not a root of unity and is rather a real number.
In this case the problem of computing $\mathcal{Z}_{K}(q)$ is \#P-hard, for which tensor contraction shows good performance \cite{Meichanetzidis2019}.

If one wishes to evaluate the Jones polynomial at non-lattice roots of unity on a quantum computer,
then one needs to work with unitary representations of the generators of the braid group \cite{Aharonov2009}.

\section{IBM Quantum Experiments and the H-test}
\label{sec:experiments-and-htest}

The experiments that we implement in this work
are for the case of \emph{Ising anyons}.
This is obtained when we set
$q=2$ in Eq.~\ref{eq:eval-point} and obtain $t=i\in\theta_\Lambda$.
Now every wire in the circuit carries one qubit
and the gates are
\begin{equation}
\mathcal{K}^\pm = \text{diag}(\pm i, 1, 1, \pm i).
\end{equation}

We consider four small knots that are all topologically equivalent under Reidemeister moves, but which give rise to different quantum circuits on different numbers of qubits:
\begin{center}
\includegraphics[width=0.6\columnwidth]{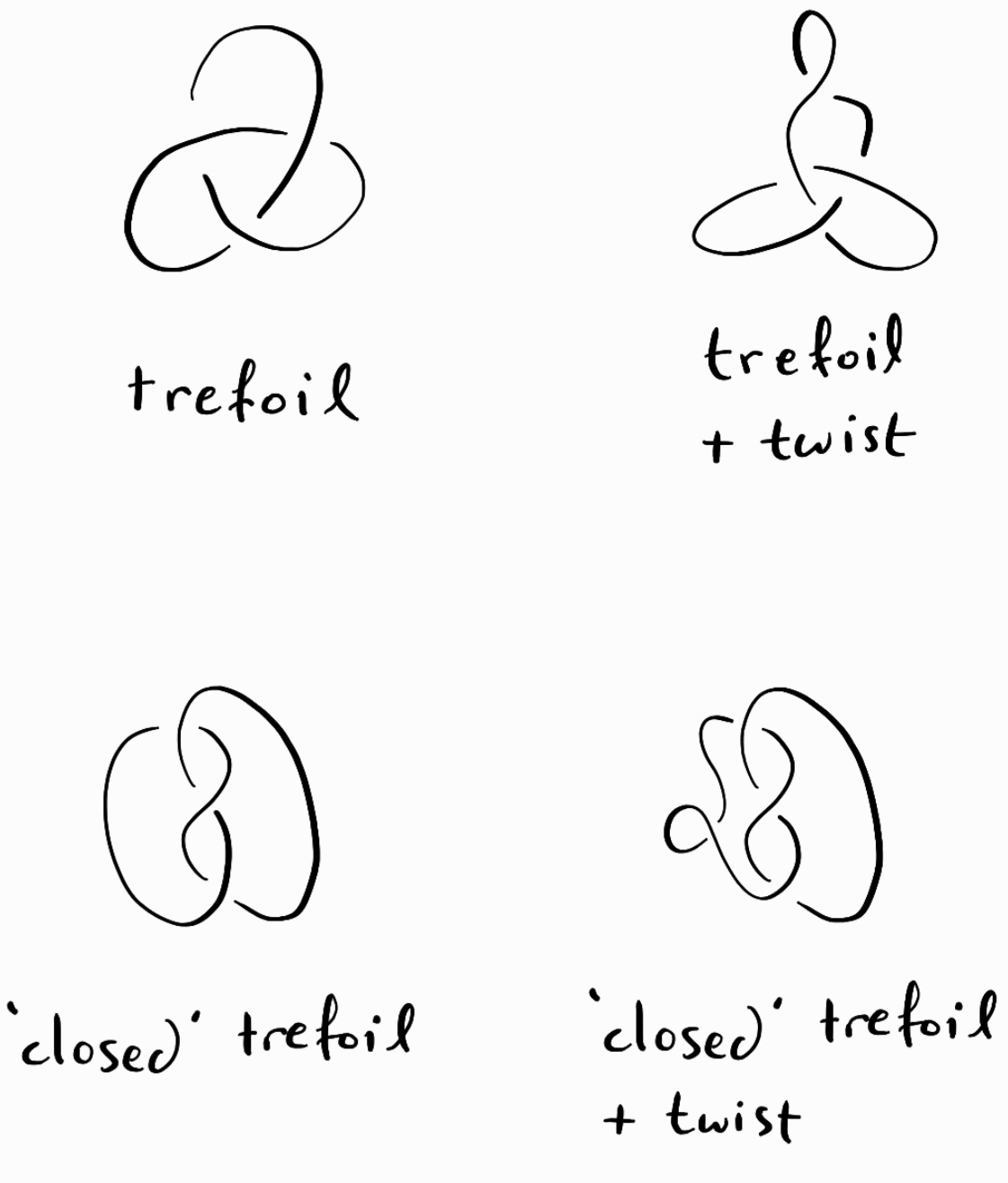}
\end{center}
The Ising anyon Jones polynomial for all of these knots $V_K(i) = -1$.
Studying four topologically equivalent knots allows us to experimentally test whether we can identify their equivalence.
Details of these knots, their qubit unitaries $C_K$ and the proportionality factors $\mathcal{A}(\tau,w,n)$, are given in appendix~\ref{app:knots}.

\begin{figure}[t]
    \centering
    \includegraphics[width=\columnwidth]{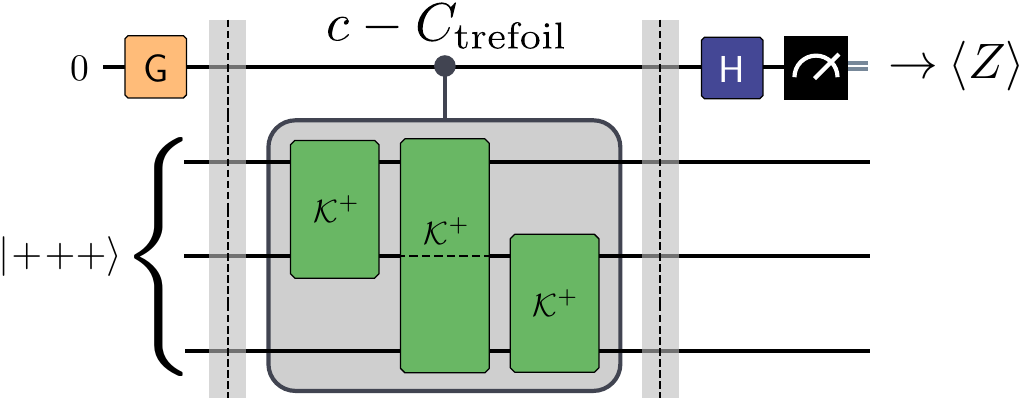}
    \caption{H-test to compute the real and imaginary part of $\mathcal{Z}_{\textrm{trefoil}} = 2^3 \langle +++ | C_\textrm{trefoil} | +++ \rangle$. Setting $G=H$ evaluates the real part $\langle Z \rangle = \textrm{Re}\{ \mathcal{Z}_{\textrm{trefoil}} \} / 2^3$, instead choosing $G = S^\dagger H$ returns the imaginary part $\langle Z \rangle = \textrm{Im} \{ \mathcal{Z}_{\textrm{trefoil}} \} / 2^3$.}
    \label{fig:trefoil-hadamard-test}
\end{figure}

Quantum amplitudes can be estimated up to additive error using the H-test,
which allows one to estimate $\langle \psi | U | \psi \rangle$
by preparing $|\phi \rangle \otimes |\psi \rangle$ on $n+1$ qubits,
and coherently controlling the unitary in question $c-U$.
Finally, the control qubit is sampled in the computational basis to estimate $\langle Z \rangle$.
Choosing $|\phi \rangle = |+ \rangle  = (| 0 \rangle + | 1 \rangle) / \sqrt{2}$ we obtain the real part of the amplitude, while setting $|\phi \rangle = (| 0 \rangle - i | 1 \rangle) / \sqrt{2}$ yields the imaginary part.
The H-test for the trefoil knot is shown in Fig.~\ref{fig:trefoil-hadamard-test}.
We compile the controlled unitary of the H-test into quantum gates by exploiting the diagonal structure of $\mathcal{K}^{\pm}$~\cite{shende2006synthesis}. This can result in redundant CNOT gates, which we remove. The circuit obtained is shown in Fig.~\ref{fig:trefoil-experimental-results}(a).
The controlled unitary does not give a Clifford circuit even though the unitary itself is Clifford, meaning the circuit contains a universal CNOT+$T$ gateset.

Experiments were carried out on IBM quantum processors to estimate the Ising anyon partition functions of the four knots, $\mathcal{Z}_{K} = 2^n \cdot {}^{\otimes n}\langle+| C_{K} |+\rangle^{\otimes n}$ where $| + \rangle$ is the normalised +1 eigenstate of the Pauli $X$ matrix. The H-test circuits were transpiled using the TKET python package~\cite{sivarajah2020t,pytket_pypi} and executed on \textit{ibmq\_lima} (quantum volume 8), \textit{ibmq\_quito} (quantum volume 16), \textit{ibmq\_paris} (quantum volume 32) and \textit{ibmq\_montreal} (quantum volume 128) backends~\cite{a_ibmq_lima,b_ibmq_quito,c_ibmq_paris,d_ibmq_montreal}. This process was repeated on different days between May and August 2021, giving between 120 and 160 separate executions on each backend including some carried out back-to-back on the same day. This is equivalent to approximately one million measurement shots in total for each backend. Additional experimental details are given in appendix~\ref{app:exp-details}.

\section{Error mitigation}
\label{sec:error-mitigation}

\begin{figure*}[t]
    \centering
    \includegraphics[width=\textwidth]{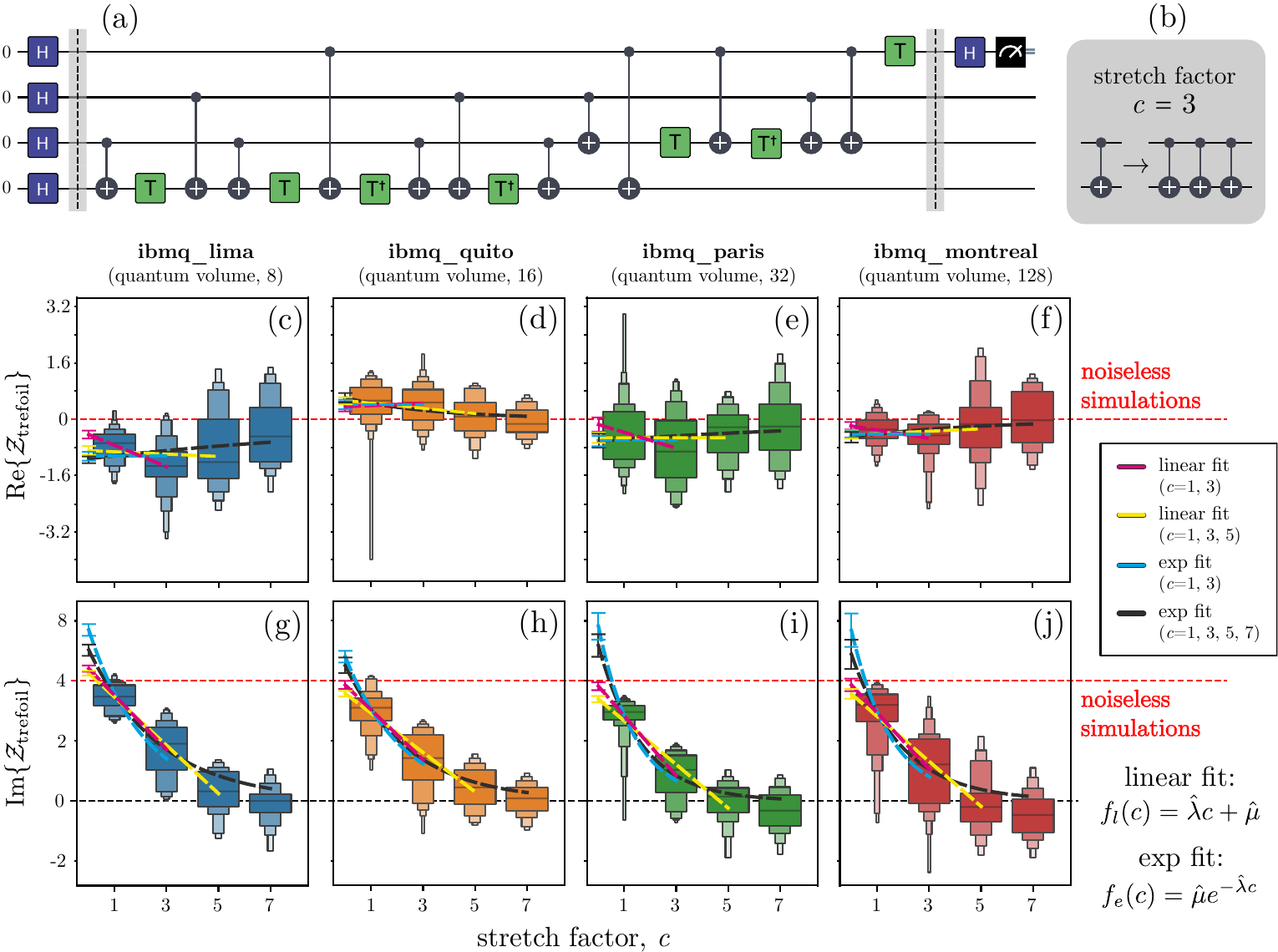}
    \caption{Experimental data for the H-test of the Ising anyon trefoil knot partition function, $\mathcal{Z}_{\textrm{trefoil}}$. (a) H-test circuit for the real part of the calculation, compiled into single qubit gates and CNOT gates. (b) Final estimates are obtained by zero-noise extrapolation. Stretch factors $c=3, 5, 7$ are implemented by replacing each CNOT gate with $c$ sequential CNOTs. (c-f) Experimental results for the real part. (g-j) Experimental results for the imaginary part. Experiments were repeatedly executed on four different IBM Quantum backends giving approximately 150 result sets for each backend, all data was collected between May and August 2021. Each panel in (c-j) shows the distribution of the data for different stretch factors (using boxen plots), as well different fits applied to the full set of data. These fits are used to obtain final estimates through extrapolation to $c=0$, we consider linear fits $f_l(c)$ and exponential decays $f_e(c)$. The values obtained from noiseless simulations are indicated on (c-j) with red dashed horizontal lines. All data shown includes measurement error mitigation.
    }
    \label{fig:trefoil-experimental-results}
\end{figure*}

Error mitigation is used to reduce the effects of noise in the quantum processor~\cite{endo2021hybrid,cirstoiu2022volumetric}. We separately mitigate against measurement errors and circuit gate errors.

Measurement errors are addressed by calibrating the measurement confusion matrix and inverting it~\cite{endo2021hybrid}. Since we only measure a single qubit in the H-test this has very little cost and we apply it to all results presented.

\begin{figure*}[t]
    \centering
    \includegraphics[width=\textwidth]{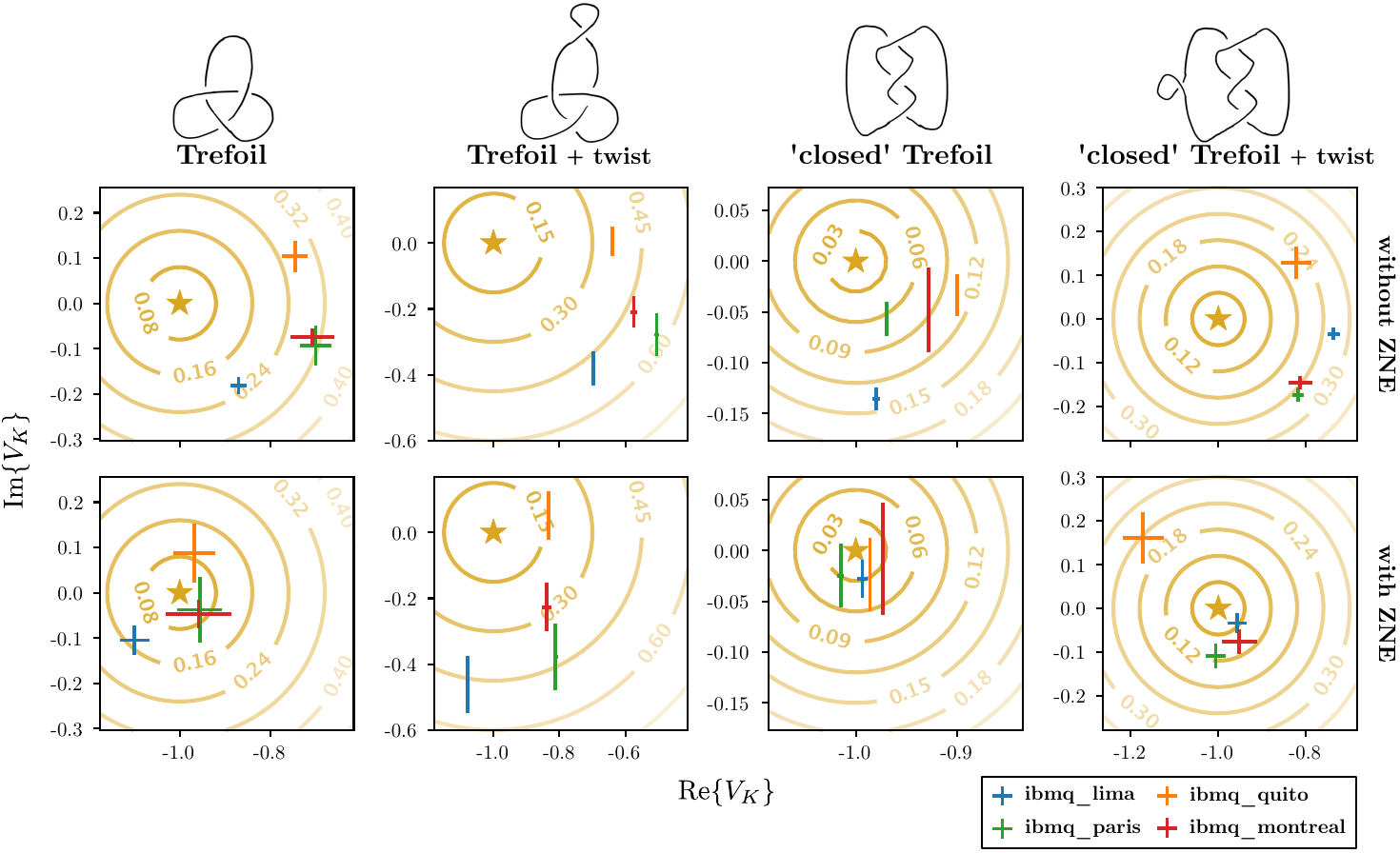}
    \caption{Jones polynomial evaluations $V_K(t=i)$ in the complex plane for four topologically equivalent knots from experiments carried out on IBM quantum hardware. Comparing (top row) results without ZNE, CNOT stretch factor $c=1$, to (bottom row) estimates obtained from zero noise extrapolation using a linear fit to $c=1, \, 3$. Each panel indicates the exact value $(-1)$ with a star, as well as showing contour lines for the distance from the exact value. The results from four different IBM Quantum devices are plotted with error bars. All results include measurement error mitigation.}
    \label{fig:jones-estimates}
\end{figure*}

To address gate errors we use zero-noise extrapolation (ZNE)~\cite{li2017efficient,temme2017error,endo2021hybrid,bharti2021noisy,cirstoiu2022volumetric}. In this approach we systematically increase the physical error rate of the gates $\lambda \rightarrow c \lambda$ for $c>1$. By measuring the target observables $\langle O \rangle$ at different stretch factors $c$ we can fit the data to a simple model $\langle O \rangle_c \sim f(c)$. ZNE estimates are obtained by extrapolating the fit to zero noise, $f(0)$. Although simple this technique has been shown to be experimentally effective~\cite{kandala2018extending,kim2021scalable,dumitrescu2018cloud}.
Focussing on the CNOT gates as the main source of errors, the error rate in the circuit is increased by replacing each CNOT in the circuit with $c$ consecutive CNOTs~\cite{dumitrescu2018cloud,giurgica2020digital,he2020resource}, Fig.~\ref{fig:trefoil-experimental-results}(b). In the noiseless case pairs of CNOTs cancel, however in the presence of noise this increases the error rate by discrete stretch factors $c = 3, 5, 7, \ldots$

Experimental data from our evaluations of $\mathcal{Z}_{\textrm{trefoil}}$ is shown as a function of ZNE stretch factor $c$ for the real part in Figs.~\ref{fig:trefoil-experimental-results}(c-f) and imaginary part in Figs.~\ref{fig:trefoil-experimental-results}(g-j). Each panel shows the distribution of the different outcomes obtained over our set of IBM Quantum evaluations as a boxen plot\footnote{Boxen plots show the distribution of data. The median is indicated in the centre. The widest two boxes directly above and below the median together contain 50\% of the data. At each decreasing level, when the boxes become narrower, the top and bottow box together contain 50\% of the remaining data. This repeats until only outliers remain.} at each $c=1,3,5,7$. The data from different days is pooled together and we fit to the whole data set, this will help average over the time varying errors resulting from drift and different calibrations. Different fits to the data are shown using either all of the $c$ or a subset, e.g. $c=1,3$. The fits we show are either linear or exponential decays towards zero. The final ZNE estimates obtained from the fits are drawn at $c=0$, along with their error bars. Results for the other three knots we consider are shown in appendix~\ref{app:exp-more-results}. 
More details of the fitting procedure and obtaining uncertainties in our ZNE estimates is given in appendix~\ref{app:fits-and-uncertainties}.

\section{Experimental results for the Jones polynomial}

We combine the evaluations of the partition function with the complex proportionality factors  $\mathcal{A}(\tau, \omega, n)$ to give the Jones polynomials $V_K(t=i)$. Fig.~\ref{fig:jones-estimates} shows the estimates of $V_K(t=i)$ obtained for the four different knots from the IBM Quantum processors. Again, we compare estimates with and without zero-noise extrapolation. For our final estimates we use the simplest fit, a linear fit to small $c=1,3$.

In most cases applying ZNE confers a clear benefit to our results. A significant improvement is realised by the smallest case we consider, `closed' trefoil ($C_K$ is a 2 qubit unitary).  However, the largest case we consider, trefoil+twist ($C_K$ is a 4 qubit unitary), shows a much weaker benefit. 
It is typically expected that error mitigation strategies will decrease the bias of a estimate while increasing its variance~\cite{endo2021hybrid} and that is the behaviour we see, with our ZNE estimates lying closer to the target value but having larger error bars.

\begin{figure}[t]
    \centering
    \includegraphics[width=0.95\columnwidth]{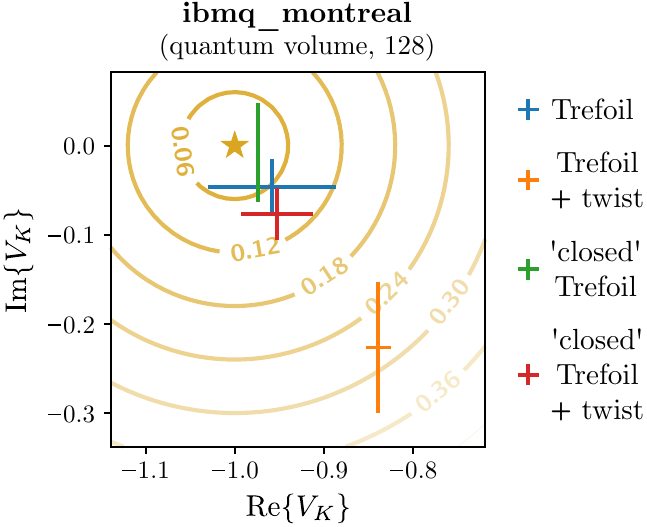}
    \caption{Comparing Jones polynomial evaluations $V_K(t=i)$ of four topologically equivalent knots obtained from the \emph{ibmq\_montreal} QPU. Values are plotted in the complex plane.
    Without noise we would expect all four knots to give the same value $V_K(t=i) = -1$, which is indicated with a star. However, errors shift the values we obtain, moving them away from $(-1)$ and causing them to not all overlap with each other. We plot contour lines showing the distance from this exact value.
    Estimates of $V_K(t=i)$ include both measurement error mitigation and zero noise extrapolation using a linear fit to $c=1, \, 3$. 
    }
    \label{fig:jones-estimates-knots-comparison}
\end{figure}

The four knots we consider are topologically equivalent meaning they all have the same Jones polynomial value $V_K(t=i) = -1$. In principle, we would expect all our experimental estimates to be consistent with each other and with $(-1)$, however, as Fig.~\ref{fig:jones-estimates} shows that is not always the case due to errors. 
To more directly make the comparison Fig.~\ref{fig:jones-estimates-knots-comparison} jointly plots the final error-mitigated estimates for the knots, evaluated on \emph{ibmq\_montreal}. We see that three of the four are consistent with each other up to error bars and are clustered together close to $(-1)$, however, the remaining knot Trefoil$+$twist is significantly apart. If we had no prior knowledge of what $V_K(t=i)$ should be we may incorrectly conclude that Trefoil$+$twist is topologically distinct from the other knots. However, if we were performing a simpler task, for instance answering whether we think this knot is more likely to be topologically equivalent to the trefoil, $V_K(t=i) = -1$, or the unknot,  $V_K(t=i) = +1$, this noisy estimate may be good enough.

\section{Related work and $\text{DQC}_1$}

The authors of Refs. \cite{PhysRevA.81.032319,onecleanqubitjonesibmq} have used the one-clean qubit model, or $\text{DQC}_1$, the `one clean qubit' model, to estimate the Jones polynomial at roots of unity, respectively using nuclear magnetic resonance technology and superconducting quantum processors provided by IBM. The latter work frames their work as a benchmark for noisy quantum computers.

In the quantum protocols used therein, one is interested in estimating the trace of a unitary matrix
by using as a resource the maximally mixed state.
The $\text{DQC}_1$ protocol is identical to the H-test,
with the difference that the maximally mixed state is used as a resource instead of a pure state.
Operationally, on the circuit model of quantum computation,
one would simulate the preparation of the maximally mixed state by averaging over H-tests
while uniformly sampling the pure state from the computational basis.

Interestingly, evaluating the Jones polynomial on roots of unity
is DQC1-complete for the trace, or `Markov', closure of a braid \cite{ShorJones},
and is BQP-complete for the closure of a braid with `cups' and `caps', called the `Plat' closure \cite{Aharonov2011,Kuperberg2015}.
These results involve unitary representations of the braid-group generators at particular roots of unity.
The braid group acts on strands that are drawn in parallel and
generators swap neighbouring strands clockwise and anticlockwise.
Their unitary representations can be seen as local unitary gates,
which when composed form arbitrary unitary transformations.
The exception to universality is for \emph{lattice roots of unity}.

One would be tempted to conclude from the above that $\text{DQC}_1=BQP$,
which is \emph{not} believed to be the case.
Alexander's theorem
states that any knot diagram $K$, which can be viewed as the Plat closure $K=P(B)$ of a braid $B$,
can be rewritten as
the Markov closure of another braid $K=M(B')$.
Moreover, the braid $B'$ can be \emph{efficiently} obtained with Vogel's algorithm \cite{BraidsSurvey2005}.
For a given root of unity,
a braid $B$ is represented as a unitary $U_B$
and the cups and caps are represented by pure basis states and basis effects \cite{KauffmanFibonacciRepresentation}.
So, $P(B)$ represents a quantum amplitude $\langle \text{cups}|U_B|\text{caps}\rangle$, which is estimated with the H-test \cite{Aharonov2009}.
Also, $M(B')$ represents $\text{tr}(U_{B'})$ and is obtained by $\text{DQC}_1$.
The subtle caveat lies in that $B'$ in general involves a larger number of strands than $B$ \cite{ShorJones}.
Furthermore, the dimension of $U_B$ depends \emph{exponentially} on the number of strands in $B$.
Since both the H-test and $\text{DQC}_1$
estimate a quantity up to additive error,
this means that, in the worst case,
for the same number of samples
the additive error in the quantum protocol involving the braid with the larger number of strands will be exponentially worse.
One can of course attempt to find a \emph{minimal braid} representative of $B'$ \cite{minimumbraid}, i.e. a braid which has the fewest possible number of strands and is topologically equivalent to $B'$ after a Markov closure, but this is a hard optimisation problem.

Note that recognising the unknot is in NP $\cap$ co-NP \cite{ComplexityKnotLink,KuperbergKnottedness2014}, and it is interestingly an open problem whether there exists an algorithm in P for it.
Remarkably, a quasipolynomial algorithm was announced for the unknotting problem recently \cite{oxMarcLackenby}.

\section{Discussion}

In this work we have rigorously demonstrated that we are able to use current quantum hardware to obtain good estimates of Jones polynomials for small knots. Further, by applying zero-noise extrapolation we were able to boost the accuracy of our estimates. 
In some cases we were able to identify the topological equivalence between the different knots we were looking at.

An obvious way to further improve our results would be to use more sophisticated error mitigation strategies.
We employed a simple approach where CNOT gates were multiplied and we made one parameter empirical fits to the data. 
Other approaches are able generate non-integer stretch the factors, allowing us to acces values of $c$ close to 1 where we would expect simple fits to the noisy observable to hold best. These methods include stretching the control pulses of the hardware in time~\cite{kandala2018extending} or randomising the insertion of extra CNOTs~\cite{he2020resource}.
Similarly, we can consider fitting models that are derived from the physical error model of the device~\cite{endo2021hybrid} or  multi-parameter scalings of the noise rates~\cite{otten2019recovering}.
Finally, the most successful applications of error mitigation combine various strategies including zero-noise extrapolation, dynamical decoupling and Pauli twirling~\cite{kim2021scalable}.

Before closing,
we recall that our setup introduces a standardised bechmark for NISQ processors.
Randomly applying Reidemeister moves to specific knots leaves the Jones polynomial invariant.
Such a procedure thus gives an efficient way to produce distributions of knots, and therefore stabiliser circuits, for which we know we should obtain the same result.
This defines a standardised benchmark for NISQ processors and error mitigation protocols.
Tangentially, our work can be generalised to the evaluation of the Jones polynomial at the sixth root of unity, in which case qutrit stabiliser circuits are involved.
Finally, one can consider evaluating the Jones polynomial
via the unitary representations of the braid group,
in which case the quantum amplitudes to which the problem reduces to involve circuits comprising gates from a universal gateset.

Finally, another general direction for further work would be to adapt this work around the problem of estimating partition functions of families of classical spin models, such as Ising or Potts.
In addition, similar benchmarks for NISQ as well as error corrected processors as what we argue for in this work can be defined in terms of classical spin models. More interestingly even, using quantum computers to execute the H-test allows for the exploration of the average case behaviour of problem families whose complexity is known, conjectured, or unknown.

\section*{Code and data availability}
Code to reproduce the experimental results presented in this paper is available from~\cite{github_link}.
In git repository \cite{knut}, connected to Ref. \cite{Meichanetzidis2019}, we provide a python script {\tt knut.py} containing functions that, given a planar code for a knot,
return the writhe, the Tait number, and the Tait graph.

The experimental data presented in this paper is available from~\cite{zenodo_link}.

\begin{acknowledgments}
KM acknowledges support from a postdoctoral fellowship by the Royal Commission for the Exhibition of 1851.
CNS acknowledges support from a Samsung GRC project and the UK Hub in Quantum Computing and Simulation, part of the UK National Quantum Technologies Programme with funding from UKRI EPSRC grant EP/T001062/1.
GKB acknowledges support from the Australian Research Council Centre of Excellence for Engineered Quantum Systems (Grant No. CE 170100009).
SI acknowledges support from Ministerio de Ciencia, Innovaci\'on y Universidades (Spain) (grant no. PGC2018-095862-B-C21-Tecnolog\'ias cu\'anticas, PID2020-113523GB-I00-An\'alisis matem\'atico y teor\'ia de informaci\'on cu\'antica, CEX2019-000918-M-Mar\'ia de Maeztu), Generalitat de Catalunya (SGR 1761), and from the European Union Regional Development Fund within the ERDF Operational Program of Catalunya (Spain) (project QUASICAT-QuantumCat 001- P-001644). 

We thank Henrik Dreyer, Matty Hoban, Yoshi Yonezawa, and Yuta Kikuchi for comments on the manuscript.

We acknowledge the use of IBM Quantum services for this work. The views expressed are those of the authors, and do not reflect the official policy or position of IBM or the IBM Quantum team.
\end{acknowledgments}


\begin{thebibliography}{10}

\bibitem{Adams2004}
Colin~C. Adams.
\newblock ``The knot book - an elementary introduction to the mathematical theory of knots''.
\newblock {American Mathematical Society}. ~(2004).

\bibitem{Kauffman2001}
Louis~H Kauffman.
\newblock ``Knots and physics''.
\newblock \href{https://dx.doi.org/10.1142/4256}{World Scientific}. ~(2001).

\bibitem{Aharonov2009}
Dorit Aharonov, Vaughan Jones, and Zeph Landau.
\newblock ``A polynomial quantum algorithm for approximating the jones polynomial''.
\newblock \href{https://dx.doi.org/10.1007/s00453-008-9168-0}{Algorithmica {\bf 55}, 395--421}~(2009).

\bibitem{endo2021hybrid}
Suguru Endo, Zhenyu Cai, Simon~C Benjamin, and Xiao Yuan.
\newblock ``Hybrid quantum-classical algorithms and quantum error mitigation''.
\newblock \href{https://dx.doi.org/10.7566/JPSJ.90.032001}{Journal of the Physical Society of Japan {\bf 90}, 032001}~(2021).

\bibitem{lubinski2023application}
Thomas Lubinski, Sonika Johri, Paul Varosy, Jeremiah Coleman, Luning Zhao, Jason Necaise, Charles~H Baldwin, Karl Mayer, and Timothy Proctor.
\newblock ``Application-oriented performance benchmarks for quantum computing''.
\newblock \href{https://dx.doi.org/10.1109/TQE.2023.3253761}{IEEE Transactions on Quantum Engineering {\bf 4}, 1--32}~(2023).

\bibitem{QCAppBenchmarks}
\url{https://github.com/SRI-International/QC-App-Oriented-Benchmarks}.

\bibitem{knotatlas}
\url{http://katlas.org}.

\bibitem{Wu}
F.~Y. Wu.
\newblock ``Knot theory and statistical mechanics''.
\newblock \href{https://dx.doi.org/10.1103/RevModPhys.64.1099}{Rev. Mod. Phys. {\bf 64}, 1099--1131}~(1992).

\bibitem{Meichanetzidis2019}
Konstantinos Meichanetzidis and Stefanos Kourtis.
\newblock ``Evaluating the jones polynomial with tensor networks''.
\newblock \href{https://dx.doi.org/10.1103/physreve.100.033303}{Physical Review E{\bf 100}}~(2019).

\bibitem{Damm2002}
Carsten Damm, Markus Holzer, and Pierre McKenzie.
\newblock ``The complexity of tensor calculus''.
\newblock \href{https://dx.doi.org/10.1007/s00037-000-0170-4}{computational complexity {\bf 11}, 54--89}~(2002).

\bibitem{Gray2021}
Johnnie Gray and Stefanos Kourtis.
\newblock ``Hyper-optimized tensor network contraction''.
\newblock \href{https://dx.doi.org/10.22331/q-2021-03-15-410}{Quantum {\bf 5}, 410}~(2021).

\bibitem{jaegervertiganwelsh1990}
F.~Jaeger, D.~L. Vertigan, and D.~J.~A. Welsh.
\newblock ``On the computational complexity of the jones and tutte polynomials''.
\newblock \href{https://dx.doi.org/10.1017/S0305004100068936}{Mathematical Proceedings of the Cambridge Philosophical Society {\bf 108}, 35–53}~(1990).

\bibitem{coecke_kissinger_2017}
Bob Coecke and Aleks Kissinger.
\newblock ``Picturing quantum processes: A first course in quantum theory and diagrammatic reasoning''.
\newblock \href{https://dx.doi.org/10.1017/9781316219317}{Cambridge University Press}. ~(2017).

\bibitem{jvdwZX}
John van~de Wetering.
\newblock ``Zx-calculus for the working quantum computer scientist''~(2020).

\bibitem{Teague}
Alex Townsend-Teague and Konstantinos Meichanetzidis.
\newblock ``Simplification strategies for the qutrit zx-calculus''~(2021).

\bibitem{Iblisdir2014}
S.~Iblisdir, M.~Cirio, O.~Boada, and G.K. Brennen.
\newblock ``Low depth quantum circuits for ising models''.
\newblock \href{https://dx.doi.org/10.1016/j.aop.2013.11.001}{Annals of Physics {\bf 340}, 205--251}~(2014).

\bibitem{gottesman}
Daniel Gottesman.
\newblock ``The heisenberg representation of quantum computers''~(1998).

\bibitem{Duncan2020}
Ross Duncan, Aleks Kissinger, Simon Perdrix, and John van~de Wetering.
\newblock ``Graph-theoretic simplification of quantum circuits with the {ZX}-calculus''.
\newblock \href{https://dx.doi.org/10.22331/q-2020-06-04-279}{Quantum {\bf 4}, 279}~(2020).

\bibitem{shende2006synthesis}
Vivek~V Shende, Stephen~S Bullock, and Igor~L Markov.
\newblock ``Synthesis of quantum-logic circuits''.
\newblock \href{https://dx.doi.org/10.1145/1120725.1120847}{IEEE Transactions on Computer-Aided Design of Integrated Circuits and Systems {\bf 25}, 1000--1010}~(2006).

\bibitem{sivarajah2020t}
Seyon Sivarajah, Silas Dilkes, Alexander Cowtan, Will Simmons, Alec Edgington, and Ross Duncan.
\newblock ``t| ket>: a retargetable compiler for nisq devices''.
\newblock \href{https://dx.doi.org/10.1088/2058-9565/ab8e92}{Quantum Science and Technology {\bf 6}, 014003}~(2020).

\bibitem{pytket_pypi}
pytket Python Package (v0.10.1), \url{https://pypi.org/project/pytket/} (2021).

\bibitem{a_ibmq_lima}
\emph{ibmq\_lima} (v1.0.9, v1.0.10, v1.0.11, v1.0.12, v1.0.13, v1.0.14), an IBM Quantum Falcon r4T Processor. IBM Quantum team. Retrieved from \url{https://quantum-computing.ibm.com} (2022).

\bibitem{b_ibmq_quito}
\emph{ibmq\_quito} (v1.0.12, v1.1.0, v1.1.1, v1.1.2, v1.1.3, v1.1.4, v1.1.5), an IBM Quantum Falcon r4T Processor. IBM Quantum team. Retrieved from \url{https://quantum-computing.ibm.com} (2022).

\bibitem{c_ibmq_paris}
\emph{ibmq\_paris} (v1.7.16, v1.7.17, v1.7.18, v1.7.19, v1.7.20), an IBM Quantum Falcon r4 Processor. IBM Quantum team. Retrieved from \url{https://quantum-computing.ibm.com} (2022).

\bibitem{d_ibmq_montreal}
\emph{ibmq\_montreal} (v1.9.11, v1.9.12, v1.9.13, v1.9.14, v1.9.15, v1.9.16, v1.10.2), an IBM Quantum Falcon r4 Processor. IBM Quantum team. Retrieved from \url{https://quantum-computing.ibm.com} (2022).

\bibitem{cirstoiu2022volumetric}
Cristina Cirstoiu, Silas Dilkes, Daniel Mills, Seyon Sivarajah, and Ross Duncan.
\newblock ``Volumetric benchmarking of error mitigation with qermit''.
\newblock \href{https://dx.doi.org/10.22331/q-2023-07-13-1059}{Quantum {\bf 7}, 1059}~(2023).

\bibitem{li2017efficient}
Ying Li and Simon~C Benjamin.
\newblock ``Efficient variational quantum simulator incorporating active error minimization''.
\newblock \href{https://dx.doi.org/10.1103/PhysRevX.7.021050}{Physical Review X {\bf 7}, 021050}~(2017).

\bibitem{temme2017error}
Kristan Temme, Sergey Bravyi, and Jay~M Gambetta.
\newblock ``Error mitigation for short-depth quantum circuits''.
\newblock \href{https://dx.doi.org/10.1103/PhysRevLett.119.180509}{Physical review letters {\bf 119}, 180509}~(2017).

\bibitem{bharti2021noisy}
Kishor Bharti, Alba Cervera-Lierta, Thi~Ha Kyaw, Tobias Haug, Sumner Alperin-Lea, Abhinav Anand, Matthias Degroote, Hermanni Heimonen, Jakob~S Kottmann, Tim Menke, et~al.
\newblock ``Noisy intermediate-scale quantum algorithms''.
\newblock \href{https://dx.doi.org/10.1103/RevModPhys.94.015004}{Reviews of Modern Physics {\bf 94}, 015004}~(2022).

\bibitem{kandala2018extending}
Abhinav Kandala, Kristan Temme, Antonio~D Corcoles, Antonio Mezzacapo, Jerry~M Chow, and Jay~M Gambetta.
\newblock ``Error mitigation extends the computational reach of a noisy quantum processor''.
\newblock \href{https://dx.doi.org/10.1038/s41586-019-1040-7}{Nature {\bf 567}, 491--495}~(2019).

\bibitem{kim2021scalable}
Youngseok Kim, Christopher~J Wood, Theodore~J Yoder, Seth~T Merkel, Jay~M Gambetta, Kristan Temme, and Abhinav Kandala.
\newblock ``Scalable error mitigation for noisy quantum circuits produces competitive expectation values''.
\newblock \href{https://dx.doi.org/10.1038/s41567-022-01914-3}{Nature Physics {\bf 19}, 752--759}~(2023).

\bibitem{dumitrescu2018cloud}
Eugene~F Dumitrescu, Alex~J McCaskey, Gaute Hagen, Gustav~R Jansen, Titus~D Morris, T~Papenbrock, Raphael~C Pooser, David~Jarvis Dean, and Pavel Lougovski.
\newblock ``Cloud quantum computing of an atomic nucleus''.
\newblock \href{https://dx.doi.org/10.1103/PhysRevLett.120.210501}{Physical review letters {\bf 120}, 210501}~(2018).

\bibitem{giurgica2020digital}
Tudor Giurgica-Tiron, Yousef Hindy, Ryan LaRose, Andrea Mari, and William~J Zeng.
\newblock ``Digital zero noise extrapolation for quantum error mitigation''.
\newblock In 2020 IEEE International Conference on Quantum Computing and Engineering (QCE).
\newblock \href{https://dx.doi.org/10.1109/QCE49297.2020.00045}{Pages 306--316}.
\newblock IEEE~(2020).

\bibitem{he2020resource}
Andre He, Benjamin Nachman, Wibe~A de~Jong, and Christian~W Bauer.
\newblock ``Zero-noise extrapolation for quantum-gate error mitigation with identity insertions''.
\newblock \href{https://dx.doi.org/10.1103/PhysRevA.102.012426}{Physical Review A {\bf 102}, 012426}~(2020).

\bibitem{PhysRevA.81.032319}
Raimund Marx, Amr Fahmy, Louis Kauffman, Samuel Lomonaco, Andreas Sp\"orl, Nikolas Pomplun, Thomas Schulte-Herbr\"uggen, John~M. Myers, and Steffen~J. Glaser.
\newblock ``Nuclear-magnetic-resonance quantum calculations of the jones polynomial''.
\newblock \href{https://dx.doi.org/10.1103/PhysRevA.81.032319}{Phys. Rev. A {\bf 81}, 032319}~(2010).

\bibitem{onecleanqubitjonesibmq}
Oktay G{\"o}kta{\c{s}}, Weng~Kian Tham, Kent Bonsma-Fisher, and Aharon Brodutch.
\newblock ``Benchmarking quantum processors with a single qubit''.
\newblock \href{https://dx.doi.org/10.1007/s11128-020-02642-4}{Quantum Information Processing {\bf 19}, 1--17}~(2020).

\bibitem{ShorJones}
Peter~W Shor and Stephen~P Jordan.
\newblock ``Estimating jones polynomials is a complete problem for one clean qubit''~(2007).

\bibitem{Aharonov2011}
Dorit Aharonov and Itai Arad.
\newblock ``The {BQP}-hardness of approximating the jones polynomial''.
\newblock \href{https://dx.doi.org/10.1088/1367-2630/13/3/035019}{New Journal of Physics {\bf 13}, 035019}~(2011).

\bibitem{Kuperberg2015}
Greg Kuperberg.
\newblock ``How hard is it to approximate the jones polynomial?''.
\newblock \href{https://dx.doi.org/10.4086/toc.2015.v011a006}{Theory of Computing {\bf 11}, 183--219}~(2015).

\bibitem{BraidsSurvey2005}
Joan~S. Birman and Tara~E. Brendle.
\newblock ``Braids''.
\newblock \href{https://dx.doi.org/10.1016/b978-044451452-3/50003-4}{Handbook of Knot TheoryPage 19–103}~(2005).

\bibitem{KauffmanFibonacciRepresentation}
Louis~H. Kauffman and Samuel~J. Lomonaco.
\newblock ``The fibonacci model and the temperley-lieb algebra''.
\newblock \href{https://dx.doi.org/10.1142/s0217979208049303}{International Journal of Modern Physics B {\bf 22}, 5065–5080}~(2008).

\bibitem{minimumbraid}
Thomas~A. Gittings.
\newblock ``Minimum braids: A complete invariant of knots and links''~(2004).

\bibitem{ComplexityKnotLink}
J.~Hass, J.C. Lagarias, and N.~Pippenger.
\newblock ``The computational complexity of knot and link problems''.
\newblock \href{https://dx.doi.org/10.1109/sfcs.1997.646106}{Proceedings 38th Annual Symposium on Foundations of Computer Science}~(1997).

\bibitem{KuperbergKnottedness2014}
Greg Kuperberg.
\newblock ``Knottedness is in np, modulo grh''.
\newblock \href{https://dx.doi.org/10.1016/j.aim.2014.01.007}{Advances in Mathematics {\bf 256}, 493–506}~(2014).

\bibitem{oxMarcLackenby}
``{M}arc {L}ackenby announces a new unknot recognition algorithm that runs in quasi-polynomial time | {M}athematical {I}nstitute --- maths.ox.ac.uk''.
\newblock \url{https://www.maths.ox.ac.uk/node/38304}.
\newblock [Accessed 18-Oct-2022].

\bibitem{otten2019recovering}
Matthew Otten and Stephen~K Gray.
\newblock ``Recovering noise-free quantum observables''.
\newblock \href{https://dx.doi.org/10.1103/PhysRevA.99.012338}{Physical Review A {\bf 99}, 012338}~(2019).

\bibitem{github_link}
Github repository \url{https://github.com/chris-n-self/Ising-anyons-Jones-polynomials-for-NISQ} (2022).

\bibitem{knut}
\url{https://gitlab.com/kourtis/tensorCSP/-/blob/master/knut.py }.

\bibitem{zenodo_link}
Zenodo repository \url{https://doi.org/10.5281/zenodo.7194971} (2022).

\end{thebibliography}


\appendix

\section{Knot specifications}
\label{app:knots}

Sections~\ref{sec:knot-diagrams}-\ref{sec:experiments-and-htest} of the main text describe how the calculation of the Jones polynomial (at $t=i$) can be performed on a quantum computer. This process begins with the knot diagram and ends with an $n$-qubit unitary, $C_K$, whose complex amplitudes ${}^{\otimes n}\langle+| C_{K} |+\rangle^{\otimes n}$  are evaluated with an H-test. When normalised, these complex amplitudes correspond to the partition function of a Potts model, $\mathcal{Z}_{K} = 2^n \cdot {}^{\otimes n}\langle+| C_{K} |+\rangle^{\otimes n}$. The Jones polynomial $V_K(t = i)$ is linked to these partition functions by a complex proportionality factor $V_K = \mathcal{A}(\tau,w,n) \mathcal{Z}_{K}$.

\begin{figure*}[t]
    \centering
    \includegraphics[width=0.7\textwidth]{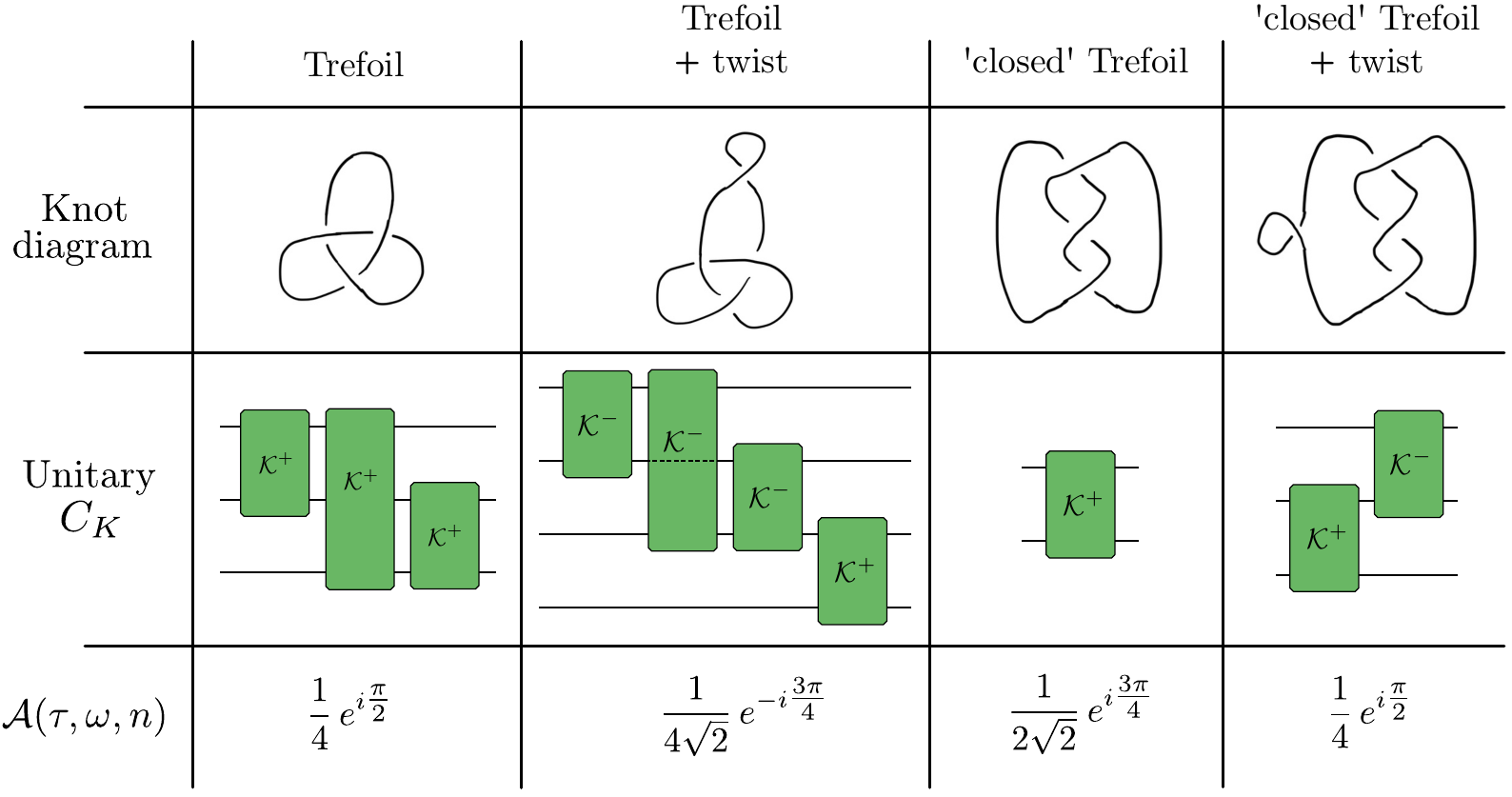}
    \caption{Specification of the knots considered on IBM quantum devixes. The four knots are topologically equivalent and each have Jones polynomial values $V_K(t=i) = -1$. For each knot, the knot diagram and related circuit unitary are shown. Additionally, the complex factors linking the circuit amplitude to the Jones polynomial, $\mathcal{A}(\tau, \omega, n)$ are given.}
    \label{fig:knots-details}
\end{figure*}

In our experiments on IBM Quantum devices we consider four knots. The knot diagrams, qubit unitaries $C_K$ and complex proportionality factors $\mathcal{A}(\tau,w,n)$ for these knots are given in Fig.~\ref{fig:knots-details}. 

The knots we consider are topologically equivalent and all have Jones polynomial $V_K(t = i) = -1$. Despite this equivalence the complexity of the circuit unitaries varies, including the number of qubits they act on. In the simplest case $C_K$ is a single two qubit unitary. This simplification occurs because of the mapping
\begin{center}
\includegraphics[scale=0.21]{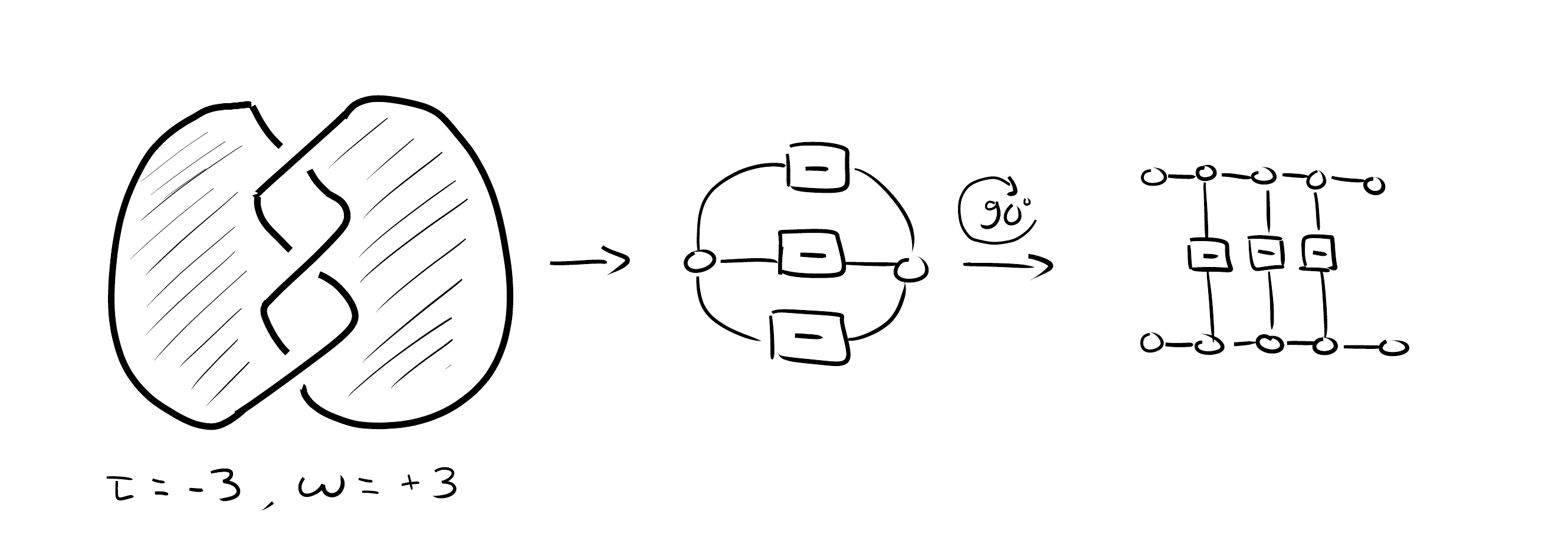}
\end{center}
followed by applying the identity
\begin{center}
\includegraphics[width=0.5\columnwidth]{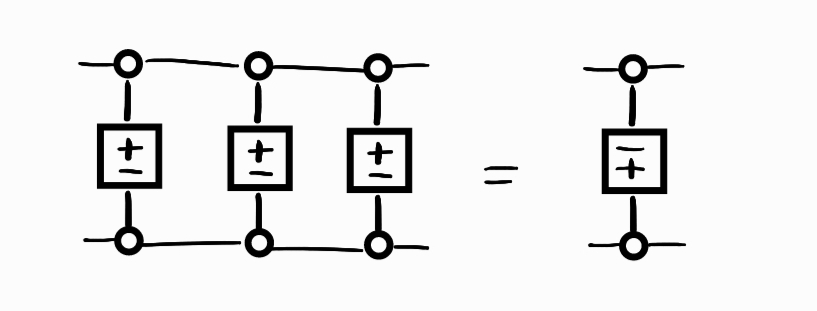}
\end{center}

\section{Experimental details}
\label{app:exp-details}

The Ising anyon partition functions, $\mathcal{Z}_{K}$, for the four knots described in Appendix~\ref{app:knots} were estimated by running experiments on IBM Quantum processors. Our experiments consisted of repeatedly submitting the quantum computation to a set of different devices over many days between May and August 2021. Here we give additional details on the practical aspects.

The controlled unitaries of the H-tests were compiled into gates using the Qiskit `diagonal' function, which implements the algorithm given in~\cite{shende2006synthesis}. 

Each day during the experiment the real and imaginary H-test circuits were submitted to the IBM Quantum backends \textit{ibmq\_lima} (quantum volume 8), \textit{ibmq\_quito} (quantum volume 16), \textit{ibmq\_paris} (quantum volume 32) and \textit{ibmq\_montreal} (quantum volume 128)~\cite{a_ibmq_lima,b_ibmq_quito,c_ibmq_paris,d_ibmq_montreal}. Before each submission the H-test circuits were transpiled for each backend using the TKET python package~\cite{sivarajah2020t,pytket_pypi} (using the default passes for IBM Quantum backends with optimisation level 2). This transpiler is noise-aware, so aspects of the transpiled circuits such as qubit selection will vary day-to-day. ZNE stretched circuits were generated from the transpiled circuit. This ensures that (1) the added CNOT's were not removed by the transpiler and (2) the ZNE stretched circuits use the same physical qubits. The circuits were batched and submitted to the IBM Quantum cloud service using the circuit submission API. Each submission consisted of four repeated copies of the real and imaginary H-test circuits at each ZNE stretch factors $c = 1, 3, 5, 7$. If the submitted circuits failed to execute before the next day the job was cancelled.

Measurement error mitigation was applied to our results using Qiskit's complete measurement fitter class. The calibration circuits were submitted batched together with our H-test circuits.

\begin{figure*}[t]
    \centering
    \includegraphics[width=\textwidth]{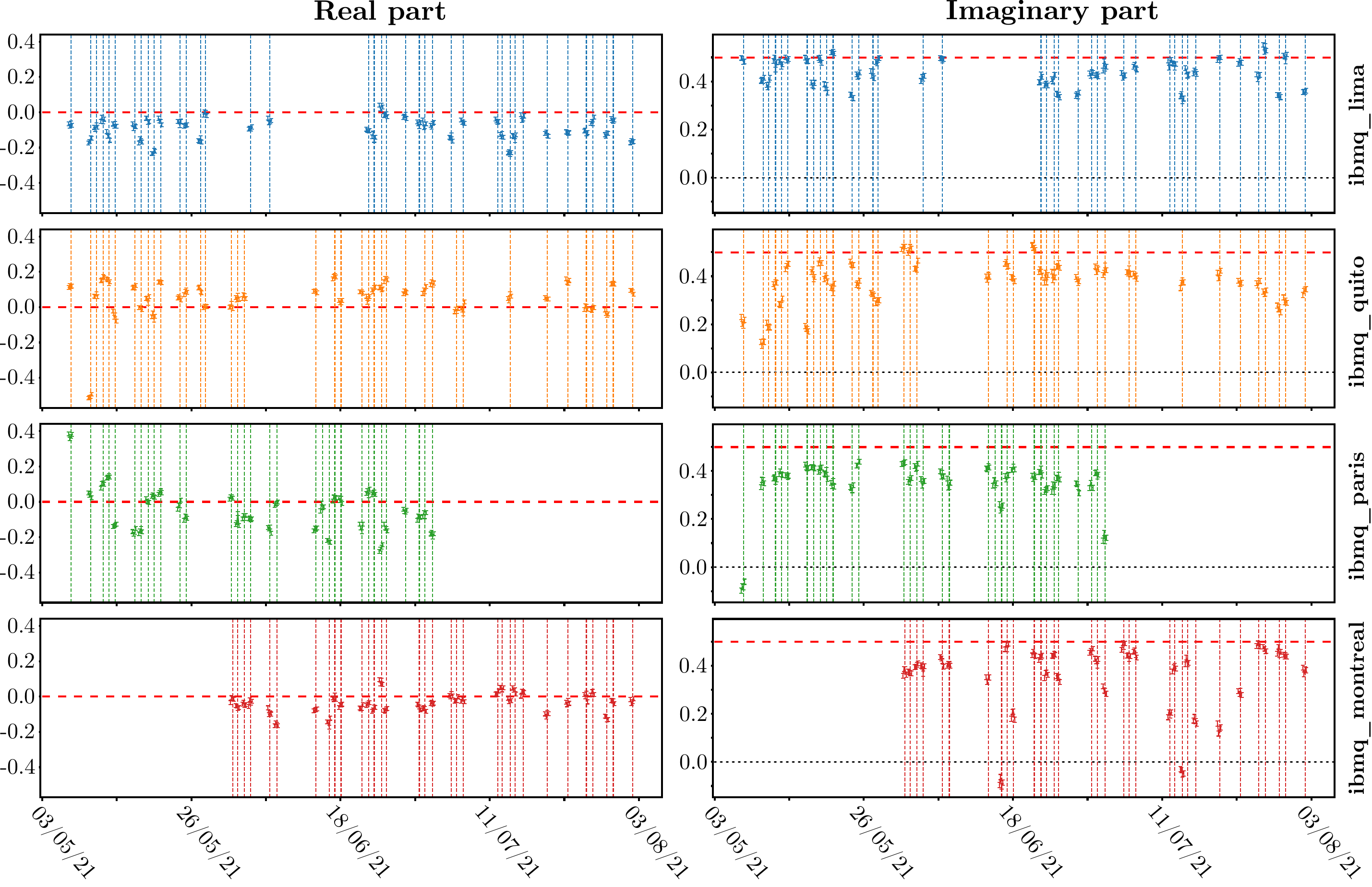}
    \caption{Timeline of evaluations of the (left) real part and (right) imaginary part of $\langle +++ | {C}_\textrm{trefoil} | +++ \rangle$ on four different IBM Quantum backends between May and August 2021. Evaluations shown are without zero-noise extrapolation ($c=1$), but after applying measurement error mitigation. Dashed vertical lines show days that we were able to collect results. Each day consisted of four batched evaluations, which are all individually plotted. All evaluations show error bars for the shot noise uncertainty. Dashed red horizontal lines show the value expected from noiseless simulations, additionally zero is highlighted with a dotted horizontal line.}
    \label{fig:trefoil-timeline}
\end{figure*}

The timeline of individual evaluations we obtained for the trefoil knot partition function is shown in Fig.~\ref{fig:trefoil-timeline}. Error bars show the uncertainty due to shot noise on each evaluation. We see that the variability between different days is much larger than the shot noise and will account for most of the uncertainty in our final estimate.

\section{Additional experimental results}
\label{app:exp-more-results}

\begin{figure*}
    \centering
    \includegraphics[width=\textwidth]{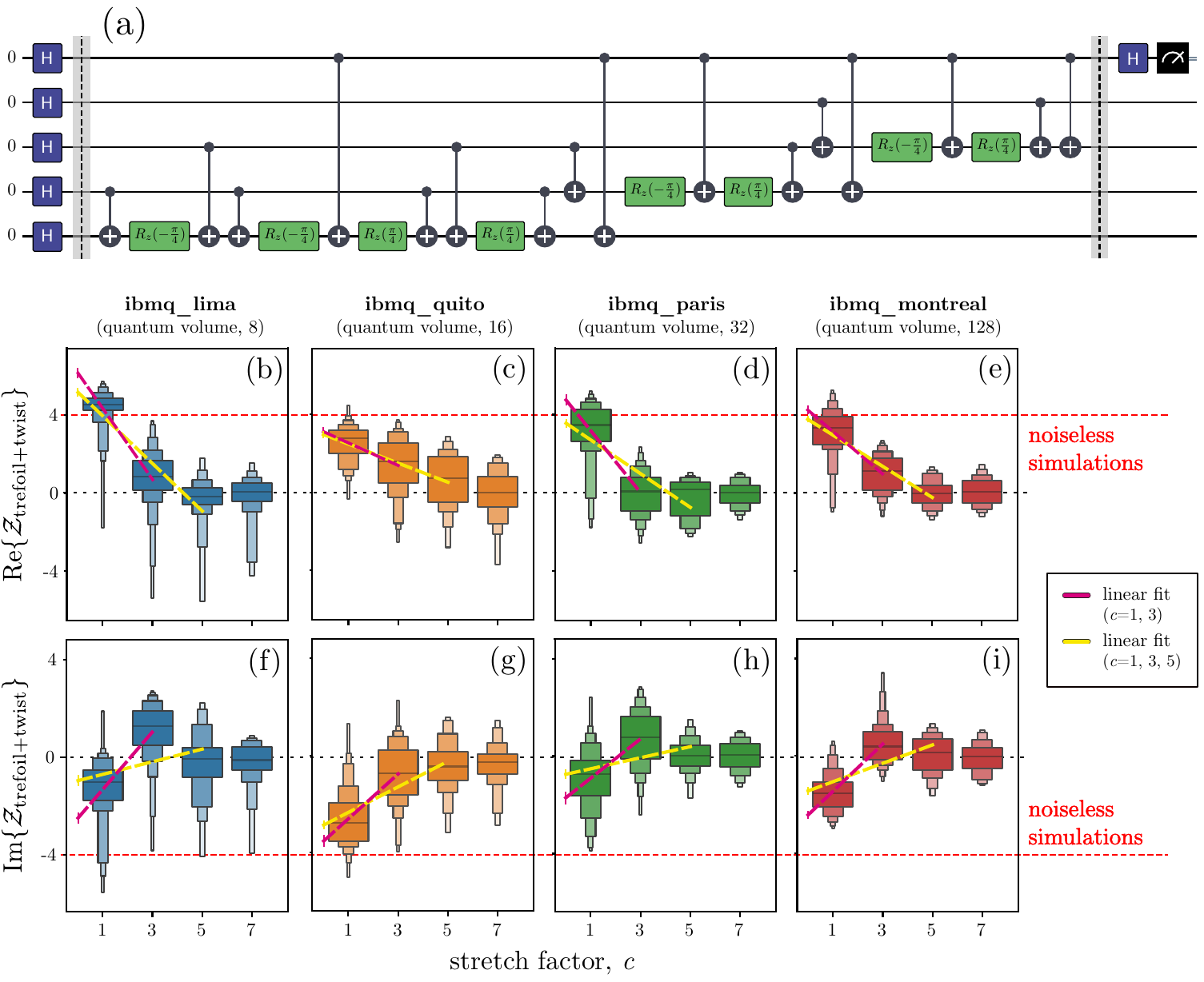}
    \caption{Experimental data for the Ising anyon trefoil with a twist knot partition function, $\mathcal{Z}_{\textrm{trefoil+twist}}$. (a) H-test circuit for the real part of the calculation, compiled into single qubit $H$, $T = R_Z(\pi/4)$ \& $T^\dagger = R_Z(-\pi/4)$ gates and two qubit CNOT gates. (b-e) Experimental results for the real part. (f-i) Experimental results for the imaginary part. Experiments were repeatedly executed on four different IBM Quantum backends giving approximately 150 result sets for each backend. Each panel in (b-i) shows the distribution of raw data for different stretch factors (boxen plots), as well different linear fits to the data. These fits are used to obtain final estimates through extrapolation to $c=0$. The values obtained from noiseless simulations are indicated on (b-i) with red dashed horizontal lines. All data shown includes measurement error mitigation.}
    \label{fig:trefoil-twist-experimental-results}
\end{figure*}

\begin{figure*}
    \centering
    \includegraphics[width=\textwidth]{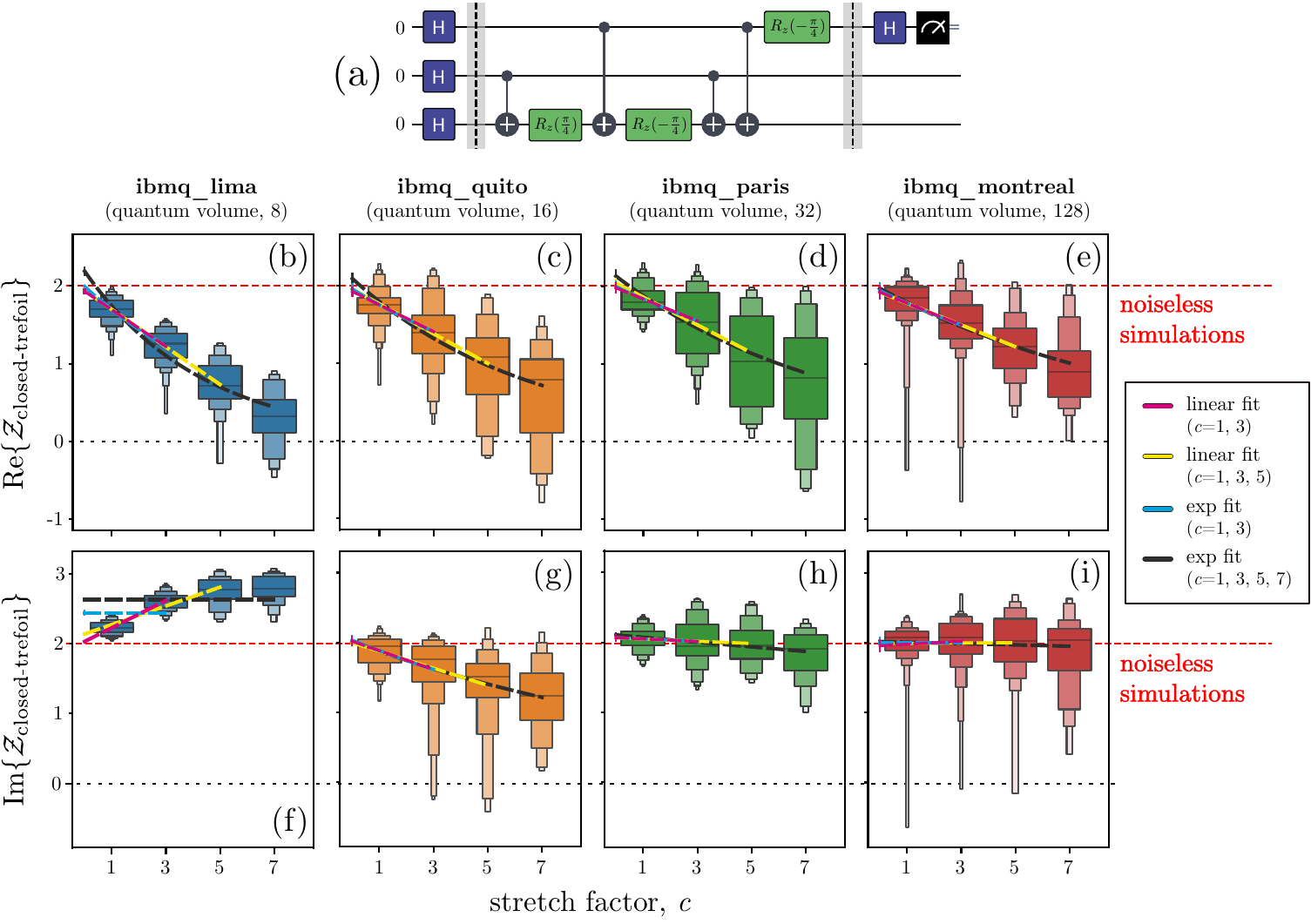}
    \caption{Experimental data for the Ising anyon `closed' trefoil knot partition function, $\mathcal{Z}_{\textrm{closed-trefoil}}$. (a) H-test circuit for the real part of the calculation, compiled into single qubit $H$, $T = R_Z(\pi/4)$ \& $T^\dagger = R_Z(-\pi/4)$ gates and two qubit CNOT gates. (b-e) Experimental results for the real part. (f-i) Experimental results for the imaginary part. Experiments were repeatedly executed on four different IBM Quantum backends giving approximately 150 result sets for each backend. Each panel in (b-i) shows the distribution of raw data for different stretch factors (boxen plots), as well different linear and exponential fits to the data. These fits are used to obtain final estimates through extrapolation to $c=0$. The values obtained from noiseless simulations are indicated on (b-i) with red dashed horizontal lines. All data shown includes measurement error mitigation.}
    \label{fig:arc-trefoil-experimental-results}
\end{figure*}

\begin{figure*}
    \centering
    \includegraphics[width=\textwidth]{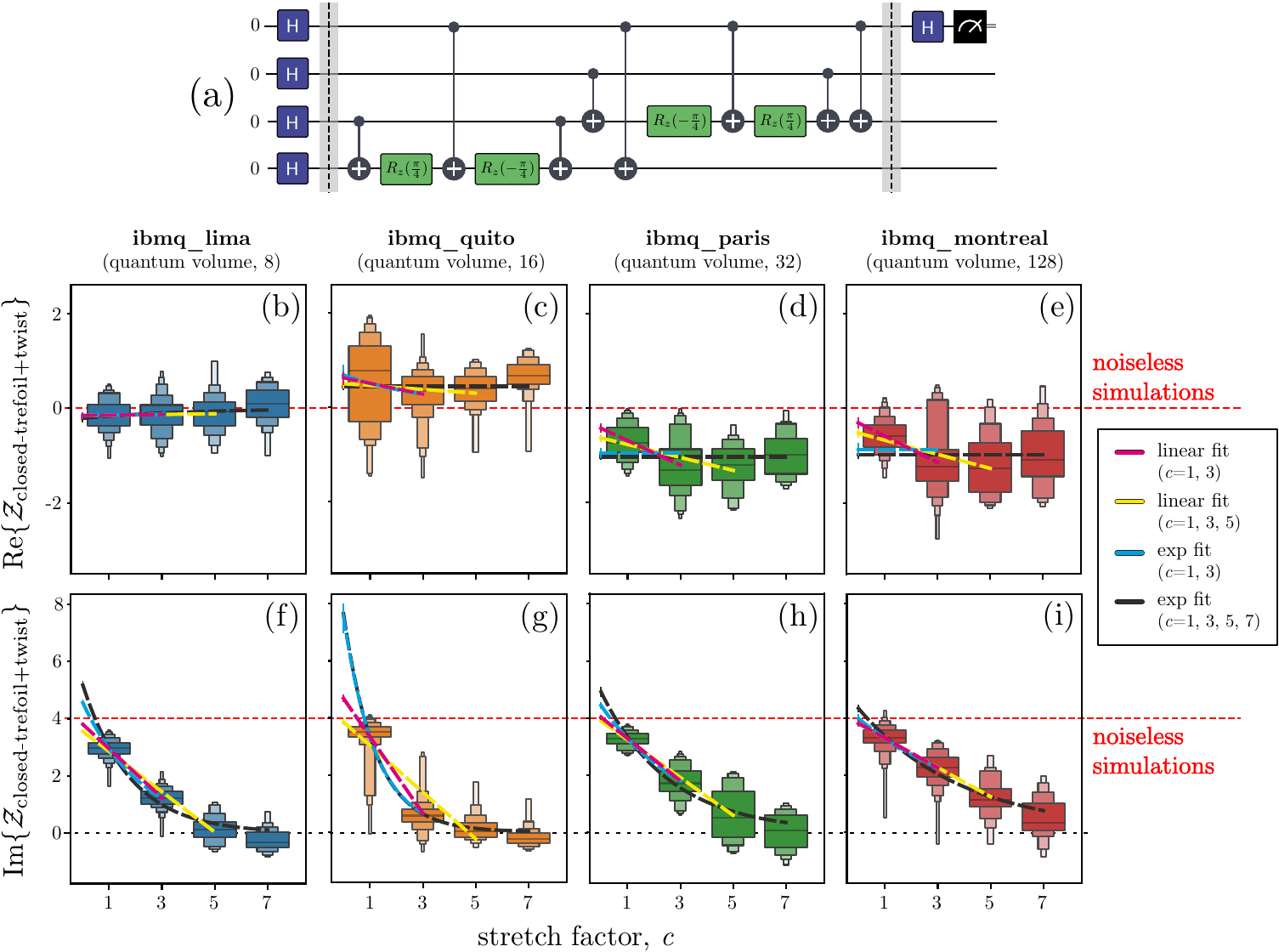}
    \caption{Experimental data for the Ising anyon `closed' trefoil with a twist knot partition function, $\mathcal{Z}_{\textrm{closed-trefoil+twist}}$. (a) H-test circuit for the real part of the calculation, compiled into single qubit $H$, $T = R_Z(\pi/4)$ \& $T^\dagger = R_Z(-\pi/4)$ gates and two qubit CNOT gates. (b-e) Experimental results for the real part. (f-i) Experimental results for the imaginary part. Experiments were repeatedly executed on four different IBM Quantum backends giving approximately 150 result sets for each backend. Each panel in (b-i) shows the distribution of raw data for different stretch factors (boxen plots), as well different linear fits to the data. These fits are used to obtain final estimates through extrapolation to $c=0$. The values obtained from noiseless simulations are indicated on (b-i) with red dashed horizontal lines. All data shown includes measurement error mitigation.}
    \label{fig:arc-trefoil-twist-experimental-results}
\end{figure*}

In section~\ref{sec:error-mitigation} of the main text we present experimental data for the H-test of the Ising anyon trefoil knot partition function. Here we show the experimental results we have for the Ising anyon partition function of the other three knots -- Fig.~\ref{fig:trefoil-twist-experimental-results} presents the data for the trefoil+twist knot, Fig.~\ref{fig:arc-trefoil-experimental-results} for the `closed' trefoil in knot and Fig.~\ref{fig:arc-trefoil-twist-experimental-results} for the `closed' trefoil+twist.
Exponential fits to the trefoil+twist data did not all converge so we only show linear fits in that case.

\section{ZNE fits and uncertainties}
\label{app:fits-and-uncertainties}

We carry out zero-noise extrapolation fits to data sets where we have pooled together repeated QPU executions. 
For concreteness consider the $c=1,3,5,7$ exponential fit to $\textrm{Re}\{ \mathcal{Z}_\textrm{trefoil} \}$. At each $c$ we have approximately 150 separate evaluations from repeated executions on different days, the exact number of repeats varies by device depending on how many job submissions were successful, for simplicity let us have exactly 150 repeats. We can label these separate estimates of $\textrm{Re}\{ \mathcal{Z}_\textrm{trefoil} \}$ as $y_{c \, t}$, for $c=1,3,5,7$ and $t=1,\ldots,150$. 
We fit the full dataset $y_{c \, t}$ to $f_e(c) = \mu e^{\lambda c}$ using the \texttt{scipy.optimise.curve\_fit} function.

Means and uncertainties for the fitting parameters are obtained by bootstrapped resampling. 
We carry out 50,000 resamples of our data $y_{c \, t}$, where at each $c$ we independently draw samples of size 150 over $t$ with repeats. Each resample is fit to $f_e(c)$ giving estimates of the fitting parameters $\mu_i$ and $\lambda_i$ for $i=1,\ldots,50000$. The mean estimates of the fitting parameters $\hat{\mu}$ and $\hat{\lambda}$ are taken as the mean of the $\mu_i$ and $\lambda_i$. Uncertainties shown on the fitting parameters in Fig.~\ref{fig:trefoil-experimental-results} are 2$\times$ the standard deviation of the $\mu_i$. These uncertainties are propagated to the Jones polynomial estimates to give the error bars in Fig.~\ref{fig:jones-estimates} and Fig.~\ref{fig:jones-estimates-knots-comparison}.
We note that independently resampling over $t$ at each $c$ scrambles experiments that were carried out at the same time. We compared it to the case where we resample tuples ($y_{1 \, t}$, $y_{3 \, t}$, $y_{5 \, t}$, $y_{7 \, t}$) over $t$, but find no difference in the results.

\clearpage
\end{document}